\documentclass[12pt]{article}
\def\hybrid{\topmargin 0pt      \oddsidemargin 0pt
        \headheight 0pt \headsep 0pt
        \voffset=-0.5cm
        \textwidth 6.25in       
        \textheight 9.5in       
        \marginparwidth 0.0in
        \parskip 5pt plus 1pt   \jot = 1.5ex}
\catcode`\@=11
\def\marginnote#1{}

\newcount\hour
\newcount\minute
\newtoks\amorpm
\hour=\time\divide\hour by60 \minute=\time{\multiply\hour by60
\global\advance\minute by-\hour}
\edef\standardtime{{\ifnum\hour<12 \global\amorpm={am}%
        \else\global\amorpm={pm}\advance\hour by-12 \fi
        \ifnum\hour=0 \hour=12 \fi
        \number\hour:\ifnum\minute<10 0\fi\number\minute\the\amorpm}}
\edef\militarytime{\number\hour:\ifnum\minute<10 0\fi\number\minute}

\def\draftlabel#1{{\@bsphack\if@filesw {\let\thepage\relax
   \xdef\@gtempa{\write\@auxout{\string
      \newlabel{#1}{{\@currentlabel}{\thepage}}}}}\@gtempa
   \if@nobreak \ifvmode\nobreak\fi\fi\fi\@esphack}
        \gdef\@eqnlabel{#1}}
\def\@eqnlabel{}
\def\@vacuum{}
\def\draftmarginnote#1{\marginpar{\raggedright\scriptsize\tt#1}}
\def\draftlabel#1{{\@bsphack\if@filesw {\let\thepage\relax
   \xdef\@gtempa{\write\@auxout{\string
      \newlabel{#1}{{\@currentlabel}{\thepage}}}}}\@gtempa
   \if@nobreak \ifvmode\nobreak\fi\fi\fi\@esphack}
        \gdef\@eqnlabel{#1}}
\def\@eqnlabel{}
\def\@vacuum{}
\def\draftmarginnote#1{\marginpar{\raggedright\scriptsize\tt#1}}

\def\draft{\oddsidemargin -.5truein
        \def\@oddfoot{\sl preliminary draft \hfil
        \rm\thepage\hfil\sl\today\quad\militarytime}
        \let\@evenfoot\@oddfoot \overfullrule 3pt
        \let\label=\draftlabel
        \let\marginnote=\draftmarginnote
   \def\@eqnnum{(\theequation)\rlap{\kern\marginparsep\tt\@eqnlabel}%
\global\let\@eqnlabel\@vacuum}  }

\def\numberbysection{\@addtoreset{equation}{section}
        \def\theequation{\thesection.\arabic{equation}}}

\def\underline#1{\relax\ifmmode\@@underline#1\else
        $\@@underline{\hbox{#1}}$\relax\fi}

\def\titlepage{\@restonecolfalse\if@twocolumn\@restonecoltrue\onecolumn
     \else \newpage \fi \thispagestyle{empty}\c@page\z@
        \def\thefootnote{\fnsymbol{footnote}} }

\def\endtitlepage{\if@restonecol\twocolumn \else  \fi
        \def\thefootnote{\arabic{footnote}}
        \setcounter{footnote}{0}}  
\relax

\numberbysection \hybrid

\newfont{\Bbb}{msbm10 scaled 1\@ptsize00}
\newfont{\Bbbb}{msbm7 scaled 1\@ptsize00}
\newcommand{\CC}{\mbox{\Bbb C}}

\newcommand{\DDD}{\raise-1pt\hbox{$\mbox{\Bbbb D}$}}

\newcommand{\PP}{\mbox{\Bbb P}}        

\newcommand{\UUU}{\raise-1pt\hbox{$\mbox{\Bbbb U}$}}

\newcommand{\ZZ}{\mbox{\Bbb Z}}
\newcommand{\z}{\raise-1pt\hbox{$\mbox{\Bbbb Z}$}}

\newcommand{\vf}{\varphi}

\newcommand{\om}{\omega}
\newcommand{\vth}{\vartheta_1}
\newtheorem{predl}{Proposition}[section]

\def\beq{\begin{equation}}
\def\eeq{\end{equation}}
\def\p{\partial}

\newtheorem{theor}{Theorem}
\newtheorem{lemma}{Lemma}[section]

\def\square{\hfill
{\vrule height6pt width6pt depth1pt} \break \vspace{.01cm}}

\begin{document}

\begin{titlepage}
\setcounter{page}{1}

\title{Classical-Quantum Correspondence \\ and\\  Functional Relations  for Painlev{\'e} Equations}

\date{}
\date{}
\date{}

\author{
A. Zabrodin
\thanks{Institute of Biochemical Physics,
Kosygina str. 4, 119991 Moscow, Russia, ITEP, Bol.
Cheremushkinskaya str. 25, 117259 Moscow, Russia and
National Research University Higher School of Economics,
20 Myasnitskaya Ulitsa, Moscow 101000, Russia,
E-mail: zabrodin@itep.ru} \and A. Zotov
\thanks{ITEP, Bol. Cheremushkinskaya str. 25, 117259 Moscow, Russia and MIPT, Inststitutskii per.  9, 141700 Dolgoprudny,
Moscow Region, Russia; E-mail: zotov@itep.ru}}

\maketitle

\vspace{-7cm} \centerline{ \hfill ITEP-TH-58/12} \vspace{7cm}

\begin{abstract}

In the light of the Quantum Painlev\'e--Calogero Correspondence established in our previous papers \cite{ZZ11,ZZ12}, we
investigate the inverse problem. We imply that this type of the correspondence (Classical--Quantum Correspondence) holds true
and find out what kind of potentials arise from the compatibility conditions of the related linear problems. The latter
conditions are written as functional equations for the potentials depending on a choice of a single function -- the
left-upper element of the Lax connection. The conditions of the Correspondence impose restrictions on this function. In
particular, it satisfies the heat equation. It is shown that all natural choices of this function (rational, hyperbolic and
elliptic) reproduce exactly the Painlev\'e list of equations. In this sense the Classical--Quantum Correspondence can be
regarded as an alternative definition of the Painlev\'e equations.

\end{abstract}


\end{titlepage}

\vfill\eject

\tableofcontents
\newpage

\section{Introduction}
\setcounter{equation}{0}

The Painlev\'e equations (${\rm P}_{\rm I}$--${\rm P}_{\rm VI}$) discovered by P.Painlev\'e, R.Fuchs and B.Gambier
\cite{Painleve1,Fuchs,Gambier} were extensively studied during the last century \cite{book,book1}. Their applications include
self-similar reductions of non-linear integrable partial differential equations \cite{FN}, correlation functions of
integrable models \cite{Bar,JMMS}, quantum gravity and string theory \cite{qg}, topological field theories \cite{Dubrovin1},
2D polymers \cite{Zamolodchikov}, random matrices \cite{TW,FW} and stochastic growth processes \cite{LTW}, conformal field
theories and KZ equations \cite{Nagoya,Iorgov}, the AGT conjecture \cite{FL,MMM} and spectral duality \cite{agt1,agt2,agt3}
to mention only few applications and few references.

As is known from classical works \cite{Fuchs,Garnier,Schlesinger} the Painlev\'e equations describe the monodromy preserving
deformations of  a system of linear differential equations with rational coefficients. The monodromy approach was developed
 by H.Flaschka, A.Newell and by M.Jimbo, T.Miwa, K.Ueno  \cite{FN, JM81-1,JM81-2,JM81-3}, see also
 \cite{IN}. At present different types of linear problems  are known (scalar \cite{Fuchs,Garnier}, 2$\times$2-matrix \cite{JM81-2} (see
also \cite{Zotov04,LOZ2}) or 3$\times$3-matrix \cite{JKT07}).

We deal with the linear problems depending on a spectral parameter \cite{Painleve1}-\cite{book1},\cite{Garnier}-\cite{IN}:
 \beq\label{d1} \left \{ \begin{array}{l}
\p_{x}\mathbf{\Psi} = {\bf U} (x,t)\mathbf{\Psi}
\\
\p_{\,t}\mathbf{\Psi} ={\bf V} (x,t)\mathbf{\Psi}
\end{array}\right. \,,
\quad \quad \mathbf{\Psi} =\left (\begin{array}{l} \psi_1
\\ \psi_{2} \end{array}\right ),
 \eeq where  ${\bf U},{\bf V}\in {sl}_2$
 explicitly depend on the spectral parameter $x$,
 on the deformation parameter $t$ (time-variable) and contain
an unknown function $u(t)$ to be constrained by the condition that the two equations have a family of common
solutions.\footnote{This function is going to satisfy one of the six Painlev\'e equations (in the Calogero form).} In fact,
the latter is equivalent to the compatibility of the linear problems expressed as the zero curvature equation (integrability
condition):
 \beq\label{P1c}
\p_x {\bf V} -\p_t {\bf U} +[{\bf V}, {\bf U}]=0\,.
 \eeq
 Set
 $$
{\bf U}=\left ( \begin{array}{cc} a & b\\  c & d
\end{array}\right ),
\quad \quad {\bf V}=\left ( \begin{array}{cc} A & B\\  C & D
\end{array}\right ).
 $$
The matrices  ${\bf U},{\bf V}$  are traceless, i. e., $a+d=0$, $A+D=0$. Then the zero curvature equation gives:
 \beq\label{zc1} \left \{ \begin{array}{l} \displaystyle{a_t -A_x +b C - c B =0}\,,
\\
\displaystyle{b_t -B_x +2aB -2bA=0}\,,
\\
\displaystyle{c_t -C_x +2cA -2a C=0\,.}
\end{array}
\right.
 \eeq

In \cite{ZZ11,ZZ12}
by applying the diagonal gauge transformation $\Omega=\hbox{diag}(\omega,\omega^{-1})$ we chose
the matrices ${\bf U}$, ${\bf V}$ such that
 \beq\label{Bb}
 b_x =2B.
 \eeq
Then the linear system (\ref{d1}) for the vector
function $\mathbf{\Psi}=(\psi_1 , \psi_2)^{\sf t}$ can be reduced to two
scalar equations for $\psi :=\psi_1$:
 \beq\label{scalareqs} \left \{
\begin{array}{l}\displaystyle{
\left (\frac{1}{2}\, \p_{x}^{2} -\frac{1}{2}\, (\p_x \log b) \, \p_x \,  + W (x,t)\right )\psi =0}\,,
\\
\displaystyle{ \p_t \psi =\left ( \frac{1}{2}\, \p_x^2 +
U(x,t)\right )\psi \,,}
\end{array}
\right.
 \eeq
where
 $$
W=U(x,t)-\frac{1}{2}\, \p_t \log b + \frac{1}{4}\, \p_{x}^2\log b+ \frac{1}{4}(\p_x \log b)^2,
 $$
and
 \beq\label{potU} U(x,t)=\frac{1}{2}\,(ad-bc -a_x) +A \, =\,
 \frac{1}{2}\, \det {\bf U}-\frac{a_x}{2}+A.
  \eeq
The second equation in (\ref{scalareqs}) has the form of a non-stationary Schr\"odinger equation in imaginary time with the
potential $U(x,t)$. It describes the isomonodromic deformations of the first one and their compatibility implies the
Painlev\'e equation (in the Calogero form) for the function $u=u(t)$:
 \beq\label{calogero} {\ddot u}=-\p_u {\tilde V}(u,t)\,,
  \eeq
generated by the Hamiltonian $H({\dot u},u,t)=\frac{1}{2}{\dot u}^2+{\tilde V}(u,t)$. The function $u=u(t)$ is defined as a
(simple) zero of the function $b(x)$:
 \beq\label{bu}
b(u)=0\,.
 \eeq

The second important condition
we are going to use together with (\ref{Bb}) is
 \beq\label{e1701}
U(x,t)=U(x,{\dot u}(t),u(t),t)=V(x,t)-H({\dot u},u,t)\,,
 \eeq
where $H({\dot u},u,t)$ is the classical Hamiltonian. The $x$-dependent part of the potential $V(x,t)$ does not contain the
dependent variable $u$. Therefore, the
second equation in (\ref{scalareqs}) acquires the form
 \beq\label{e1708}
\p_t \Psi(x,t) =\left ( \frac{1}{2}\, \p_x^2 +V(x,t)\right )\Psi(x,t)
 \eeq
with
 \beq\label{e1709}
 \Psi(x,t)=e^{\int^t H (\dot u , u,t')dt'}\psi (x,t)\,.
 \eeq
Notice that condition (\ref{Bb}) can be easily satisfied by
choosing a suitable gauge.
However, together with (\ref{e1701}) it becomes a non-trivial condition
and leads to the {\em Quantum Painlev{\'e}-Calogero Correspondence}
(see below) which relates the potentials of the classical
problem $\tilde V$ with $V$ in the quantum one. It appears that the potentials differ only by ``quantum corrections'' of
the coupling constants. Therefore, (\ref{e1708}) is the quantization of (\ref{calogero}) with the unit Planck constant.

In \cite{ZZ11,ZZ12} it was shown that there
exists a choice of gauge and variables $(x,t)$ such that the
non-stationary Schr\"odinger equation becomes a quantized
Painlev\'e equation. Thus, the linear problem
(\ref{d1}) leads to both classical and quantum Painlev\'e equations. The classical one is written in the variable $u(t)$ and
follows from the zero-curvature equation (\ref{d1})
valid for all $x$. The quantum one is written in terms of the spectral
parameter $x$ for a component of the common solution $\psi_1$ of the linear problems. We have called this construction the {\em Quantum
Painlev\'e-Calogero Correspondence}. It is a quantum version of the classical correspondence introduced by A. Levin and M. Olshanetsky \cite{LO97} and developed by K. Takasaki \cite{Takasaki01}. It should be mentioned that the phenomenon similar to
the Quantum Painlev\'e-Calogero Correspondence \cite{ZZ11,ZZ12} was first observed by B.Suleimanov \cite{Suleimanov94,Suleimanov08}
in terms of rational linear problems.

Let us note that the phenomenon of the
classical-quantum correspondence is also known in the theory
of integrable systems in some other contexts.
There are interrelations between classical and quantum
problems of a simingly different type
\cite{Zabrodin1,Zabrodin2,Zabrodin3}, where
Bethe vectors of integrable quantum spin chains are erlated to some
data of classical integrable many-body systems.
A similarity between quantum transfer matrices and
classical $\tau$-functions was pointed out in \cite{9,10,11}.

{\bf The aim of this paper} is to address
the inverse problem. We start with the system of
scalar equations (\ref{scalareqs}) and assume
that the Quantum Painlev\'e-Calogero
Correspondence takes place, i. e.,
equations (\ref{calogero})-(\ref{e1701}) hold true
(in this paper we refer to it as {\em Classical-Quantum Correspondence}
since it is not clear initially which equations
satisfy the conditions). Then we derive and solve functional equations\footnote{It should be mentioned that
functional equations play a very important
role in the theory of integrable systems; they underlie
the Lax equations, the $r$-matrix and other
structures \cite{Krich,Func}.} for
the potential $V$ searching through possible
choices of the function $b$. In other words,
we assume that the Classical-Quantum
Correspondence holds true and find out
what kind of potentials arise from the compatibility conditions.

We prove the following
 \begin{theor}
Let the compatibility condition for the system (\ref{scalareqs})
with $$U(x,{\dot u}(t),u(t),t)=V(x,t)-H({\dot u},u,t)$$ and
$$H({\dot u},u,t)=\frac{1}{2}{\dot u}^2+{\tilde V}(u,t)$$ be equivalent to
 $$
{\ddot u}=-\p_u {\tilde V}(u,t)\,,
 $$
where $u$ is defined as a simple zero of the function $b(x,t)$: $b(x,t)\left.\right|_{x=u}=0$. Then there are two
possibilities:

{\bf 1.}
 \beq\label{e2203}
 b(x,u,t)=b(x-u,t)\,.
 \eeq
The function $b(z,t)$ satisfies the heat equation
 \beq\label{e2204}
 2\p_t b(z,t)=\p_z^2b(z,t)\,,
 \eeq
the quantum potential coincides with the classical one,
 \beq\label{e2205}
 \tilde{V}(u,t)=V(u,t),
 \eeq
and satisfies the following functional equation:
 \beq\label{e755}
  \begin{array}{|c|}
  \hline\\
 V_t(x)-V_t(u)-\frac{1}{2}f(x\!-\!u,t)\bigl ( V'(x)+V'(u)\bigr )-
 f_x(x\!-\!u, t)\left(V(x)-V(u)\right)=0\,.
 \\ \ \\ \hline
 \end{array}
 \eeq
where $f(x,t)=b_x(x,t)/b(x,t)$.

{\bf 2.}
 \beq\label{e2303}
 b(x,u,t)=b(x-u,t)b(x+u,t)\,.
 \eeq
The function $b(z,t)$ satisfies the heat equation
 \beq\label{e2304}
 2\p_t b(z,t)=\p_z^2b(z,t)\,,
 \eeq
the classical and quantum potentials are related by
 \beq\label{e2305}
 \left. \phantom{\int}
 \tilde{V}(u,t)=V(u,t)+\frac{1}{2}\, \p_x^2 \log b(x,t)\right |_{x=2u}
 \eeq
and $V(x,t)$ satisfies the following functional equation:
  \beq\label{e955}
 \begin{array}{|c|}
   \hline\\
 V_t(x)-V_t(u)
 -\frac{1}{2} f(x\!-\!u,t)\left(V'(x)+V'(u)\right)
 -\frac{1}{2}f(x\!+\!u,t)
 \left(V'(x)-V'(u)\right)\\ \ \\
+\left(f_x(x\!-\!u,t)+
f_x(x\!+\!u,t)\right)
 \left(V(u)-V(x)\right)=0\,.\\ \ \\ \hline
 \end{array}
 \eeq
 \end{theor}
where $f(x,t)=b_x(x,t)/b(x,t)$.
The proof of the Theorem is based on Propositions
\ref{prop0}, \ref{prop1} and \ref{prop2}.

Solving  equations (\ref{e755}) and (\ref{e955})
we get the following results: for the rational (in $x$)
function $b$ we obtain ${\rm
P}_{\rm I}$, ${\rm P}_{\rm II}$
from (\ref{e755}) and ${\rm P}_{\rm IV}$
from (\ref{e955}), for the hyperbolic we obtain ${\rm
P}_{\rm III}$ from (\ref{e755}) and ${\rm P}_{\rm V}$
from (\ref{e955}). The most general
equation ${\rm P}_{\rm VI}$ arises for the
$\theta$-functional ansatz for $b$ from (\ref{e955}) while
the equation from (\ref{e755}) is shown to have
only trivial solutions in this case.

Finally, it is shown that all natural choices of
the function $b$
(rational, hyperbolic and elliptic) reproduce exactly the
Painlev\'e list of equations. In this sense the Classical-Quantum
Correspondence can be viewed as an alternative
definition for the Painlev\'e equations.

The paper is organized as follows. In the next section we recall the Quantum Painlev\'e-Calogero Correspondence.
In Section 3 we derive the functional equations from (\ref{e755})
and (\ref{e955}) and then solve these equations in Sections 4-6. In the appendices we give the definitions and identities for
necessary elliptic functions, discuss some special cases of the $b$-function and list the ${\bf U}$-${\bf V}$ pairs for ${\rm
P}_{\rm I}$-${\rm P}_{\rm V}$ which are acceptable for the Quantum Painlev\'e-Calogero Correspondence.

\section{Quantum Painlev{\'e}-Calogero Correspondence}
\setcounter{equation}{0}

In \cite{ZZ11,ZZ12} we described the Quantum Painlev{\'e}-Calogero Correspondence which states that for Painlev\'e equations
the non-stationary Baxter equation at $\hbar=1$ represents  a classical linear problem. Let us start from example.


\subsection{Example of Painlev\'e V}


The ${\rm P}_{\rm V}$ equation is conventionally written as:
 \beq\label{P5} \p^2_{T} y=\left (\frac{1}{2y}+\frac{1}{y\!-\! 1}\right )
(\p_{T} y)^2 -\frac{\p_{T} y}{T} +\frac{y(y\!-\! 1)^2}{T^2} \left (\alpha +\frac{\beta}{y^2}+\frac{\gamma T}{(y\!-\! 1)^2}
+\frac{\delta T^2 (y+1)}{(y\!-\! 1)^3}\right ),
 \eeq
 where $\alpha$, $\beta , \gamma , \delta$ are  parameters\footnote{There are in fact three essentially independent
parameters.}. Making change of variable
 \beq\label{P5aa} y=\coth ^2 u \eeq
together with
 \beq\label{P5aa144} T =e^{2t} \eeq
 ${\rm P}_{\rm V}$
acquires the form
 \beq\label{P5aaa} \ddot u
=-\frac{2\alpha \cosh u}{\sinh^3 u}- \frac{2\beta \sinh u}{\cosh^3 u} -\gamma e^{2t}\sinh (2u) -\frac{1}{2}\, \delta
e^{4t}\sinh (4u)\,. \eeq
 The later equation is Hamiltonian with $H_{\rm V}(p, x)=\frac{p^2}{2}+V_{\rm V}(u,t)$, where
 \beq\label{P5aaab} V_{\rm V}(u,t)=-\, \frac{\alpha}{\sinh^2 u} -\, \frac{\beta}{\cosh^2
u}+\frac{\gamma e^{2t}}{2} \cosh (2u)+\frac{\delta e^{4t}}{8}\cosh (4u). \eeq

The zero curvature representation is known from \cite{JM81-2}. It is rational in spectral parameter $X$. As it was shown in
\cite{ZZ11} the change
 \beq\label{p1} X=\cosh ^2 x \eeq
with (\ref{P5aa}) and (\ref{P5aa144}) and some special gauge transformation brings the Jimbo-Miwa ${\bf U}-{\bf V}$ pair to
the one given in (\ref{la8})-(\ref{la11}). Then the first component of the linear problem (\ref{d1}) $\psi$ satisfies the
non-stationary Schr\"odinger equation
 \beq\label{P5f} \p_t \psi =\Bigl (H_{\rm V}^{(\alpha -\frac{1}{8}, \, \beta
+\frac{1}{8}, \, \gamma ,\, \frac{1}{2} )}(\p_x , x)- H_{\rm V}^{(\alpha , \beta , \gamma , \frac{1}{2} )}(\dot u , u )\Bigr
)\psi
 \eeq
and, therefore,
 \beq\label{P5j} \p_t \Psi = H_{\rm V}^{(\alpha -\frac{1}{8}, \, \beta +\frac{1}{8}, \, \gamma ,\, \frac{1}{2}
)}(\p_x , x)\Psi =\left ( \frac{1}{2}\, \p_x^2 + V_{\rm V}^{(\alpha -\frac{1}{8}, \, \beta +\frac{1}{8}, \, \gamma ,\,
\frac{1}{2} )}(x,t)\right )\Psi
 \eeq
for $\Psi (x,t)=e^{\int^t H_{\rm V}^{(\alpha , \, \beta , \, \gamma ,\, \frac{1}{2} )} (\dot u , u)dt'}\psi (x,t)$, i. e. the
linear problem admits the form of the quantized equation (in spectral parameter). Notice that the parameters $\alpha$,
$\beta$ are shifted by $\pm \frac{1}{8}$ in the quantum Hamiltonian.

\subsection{Summary}
The following theorem summarizes the results of \cite{ZZ11},\cite{ZZ12} for all Painlev\'e equations, see also the table of
changes of variables below.

\noindent
\paragraph{Theorem \cite{ZZ11,ZZ12}}{\it  For any of the six equations
from the Painlev\'e list written in the Calogero form
as classical non-autonomous Hamiltonian systems with time-dependent Hamiltonians $H(p,u,t)$
there exists a pair of compatible linear problems
 \beq\label{d001} \left \{ \begin{array}{l} \p_{x}\mathbf{\Psi} ={\bf U}
(x,t,u,\dot{u},\{c_k\})\mathbf{\Psi}
\\
\p_{t}\mathbf{\Psi} ={\bf V} (x,t,u,\dot{u},\{c_k\})\mathbf{\Psi}
\end{array}\right. \,,
\quad \quad \mathbf{\Psi} =\left (\begin{array}{l} \psi_1 \\
\psi_{2} \end{array}\right ),
 \eeq
 where ${\bf U}$ and ${\bf V}$ are $sl_2$-valued functions, $x$ is a spectral parameter,
$t$ is the time variable and $\{c_k\}=\{\alpha , \beta , \gamma , \delta \}$ is the set of parameters involved in the
Painlev\'e equation, such that
 \begin{itemize}
\item[1)]
The zero curvature condition \beq\label{d002}
\p_t{\bf U}-\p_x {\bf V}+[{\bf U},{\bf V}]=0 \eeq
is equivalent to the
Painlev\'e equation
for the variable $u$ defined as any (simple) zero of the right
upper element of the matrix ${\bf U}(x,t)$ in the spectral
parameter: ${\bf U}_{12}(u, t)=0$;
\item[2)]
The function $\Psi =e^{\int^t H(\dot u, u,t')dt'}\psi _1$
where $\psi_1$ is the first component of $\mathbf{\Psi}$ satisfies
the non-stationary Schr\"odinger equation in imaginary time
 \beq\label{d003} \p_t\Psi =\left(\frac{1}{2}\, \p_x^2 +\tilde V
(x,t) \right)\Psi \eeq with the potential
 \beq\label{d004}
\tilde V(x,t)=V(x,t, \{\tilde c_k\})\,,
 \eeq
  \beq\label{d005}
V(x,t, \{\tilde c_k\})-\left(\frac{1}{2}{\dot u}^2+V(u,t, \{c_k\})\right)= \frac{1}{2}\Bigl [\det({\bf U})-\p_x {\bf
U}_{11}+2{\bf V}_{11}\Bigr ]\,,
 \eeq
which coincides with the classical potential $V(u,t)=V(u,t, \{c_k\})$ up to possible shifts of the parameters $\{c_k\}$:
  \beq\label{d0051}
\begin{array}{l}
(\tilde \alpha , \tilde \beta  ) =(\alpha , \beta +\frac{1}{2})
\quad \mbox{for ${\rm P}_{\rm IV}$},
\\
(\tilde \alpha , \tilde \beta , \tilde \gamma , \tilde \delta )= (\alpha -\frac{1}{8}, \beta +\frac{1}{8}, \gamma , \delta )
\quad \mbox{for ${\rm P}_{\rm V}$}
\\
(\tilde \alpha , \tilde \beta , \tilde \gamma , \tilde \delta )= (\alpha -\frac{1}{8}, \beta +\frac{1}{8}, \gamma
-\frac{1}{8}, \delta +\frac{1}{8}) \quad \mbox{for ${\rm P}_{\rm VI}$}.
\end{array}
 \eeq
 \end{itemize}
} The list of changes of variables is summarized in the following table:
 \begin{center}
\begin{tabular}{|c|c|c|c|c|}
\hline $\displaystyle{\phantom{\int}}$ Equation $\displaystyle{\phantom{\int_{\int}^{\int}}}$
& $y(u,t)$ & $T(t)$ & $X(x,t)$ & ${\bf U}_{12}(x,t)$ \\
\hline $\displaystyle{\phantom{\int}}$ ${\rm P}_{\rm I}$
& $u$ & $t$ & $x$ &$x-u$\\
\hline $\displaystyle{\phantom{\int}}$ ${\rm P}_{\rm II}$
& $u$ & $t$ & $x$ &$x-u$\\
\hline $\displaystyle{\phantom{\int}}$ ${\rm P}_{\rm IV}$
& $u^2$ & $t$ & $x^2$ &$x^2 -u^2$ \\
\hline $\displaystyle{\phantom{\int}}$ ${\rm P}_{\rm III}$
& $e^{2u}$ & $e^t$ & $e^{2x}$ & $2e^{t/2}\sinh (x-u)$ \\
\hline $\displaystyle{\phantom{\int}}$ ${\rm P}_{\rm V}$ & $\coth ^2 u$ & $e^{2t}$ & $\cosh ^2 x$ & $2e^{t}
\sinh (x\! -\! u)\sinh (x\! +\! u)$ \\
\hline $\displaystyle{\phantom{\int^{A}_{A}}} {\rm P}_{\rm VI}$ & $ \frac{\wp (u)-\wp (\omega _1)}{\wp (\omega_2)-\wp
(\omega_1)}$ & $ \frac{\wp (\omega _3)-\wp (\omega _1)}{\wp (\omega_2)-\wp (\omega_1)}$ & $ \frac{\wp (x)-\wp (\omega
_1)}{\wp (\omega_2)-\wp (\omega_1)} $ & $
\vartheta _1(x-u)\vartheta _1(x+u) h(u,t) $ \\
\hline
\end{tabular}
 \end{center}

\noindent Function $h(u,t)$ for ${\rm P}_{\rm VI}$ case can be found in \cite{ZZ12}. Notice that the given above changes of
variables can be derived in a general form from (\ref{Bb}) and requirement that the potential (\ref{potU}) could be presented
as a sum of two parts depending on $x,t$ and $u,t$ separately. This calculation was made in \cite{ZZ12} for the most general
- Painlev\'e VI equation.  The appropriate ${\bf U}$-${\bf V}$ pairs  for ${\rm P}_{\rm I}$-${\rm P}_{\rm V}$ are given in
the Appendix C.

\section{The Scalar Linear Problems and Functional Equations}
\setcounter{equation}{0}

It was shown in \cite{ZZ11,ZZ12} that each of the six equations from the Painlev\'e list, hereinafter referred to as ${\rm
P}_{\rm I}$ -- ${\rm P}_{\rm VI}$, written in the so-called
Calogero form, can be obtained as integrability conditions for
two Schr\"odinger-like equations
 \beq\label{W0}
\left \{
\begin{array}{l}\displaystyle{
\left (\frac{1}{2}\, \p_{x}^{2} -\frac{b_x}{2b}\,
\p_x \,  + W (x,t)\right )\Psi =0}
\\ \\
\displaystyle{
\p_t \Psi =\left ( \frac{1}{2}\, \p_x^2 +
V(x,t)\right )\Psi \,,}
\end{array}
\right.
 \eeq
stationary and non-stationary. The time-dependent potentials $W$ and $V$ are related by
 \beq\label{W01}
W(x,t)={U}-\frac{2{\dot b}-b_{xx}}{4b}=V(x,t)-H- \frac{2{\dot b}-b_{xx}}{4b}
 \eeq
where $H$ does not depend on $x$ and $b$ is some function of the spectral parameter $x$ and time $t$ to be chosen in such a
way that the two linear problems be compatible for some $V(x,t)$. Suppose it has a (simple) zero at the point $x=u=u(t)$:
$b(u,t)=0$ and let $V(x,t)$ be a function that depends on $x,t$ in an explicit way only (i. e., $V(x,t)$ does not contain
$u$). Let also $H$ be a function of $u$ and $\dot u$.

\noindent {\bf Remark.} Note that function $b$ may depend on $t$ in two ways -- explicit and implicit.
 The latter means
the time dependence through the unknown functions of $t$ (dependent variables). Writing $\p_t b$ we mean the derivative with respect to
the explicit dependence only. For example, $\p_t(z-u)=0$. The lower index $t$ means the same ($\p_t b(z,u(t),t)=b_t$) while the
dot is the full time derivative: ${\dot b}(z,u(t),t)={\dot u}\p_u b+b_t$. The same notations are used for other functions
depending on $t$ and $u(t)$ apart from $\Psi$ in the linear problem
(where the partial derivative symbols $\p_x$, $\p_t$
are traditionally used but, in fact,
the operator $\p_t$ acts as the full time-derivative).

Combining equations (\ref{W0}), one can write another pair
of linear problems whose compatibility implies the Painlev\'e
equations:
 \beq\label{W03}
\left \{
\begin{array}{l}\displaystyle{
\left (\frac{1}{2}\, \p_{x}^{2} -\frac{b_x}{2b}\,
\p_x \,  + W (x,t)\right )\Psi =0}
\\ \\
\displaystyle{ \p_t \Psi =\left (\frac{b_x}{2b}\, \p_x +\frac{2{\dot b}-b_{xx}}{4b}+H \right )\Psi \,,}
\end{array}
\right.
 \eeq (The first equation is the same while the other one is
a first order equation.) Passing to the function $\tilde \Psi =\Psi
/\sqrt{b}$, we can write these linear problems in the Fuchs-Garnier
form:
 \beq\label{FG1}
\left \{
\begin{array}{l}\displaystyle{
\left (\frac{1}{2}\, \p_{x}^{2}  + S(x,t)\right )\tilde \Psi =0}
\\ \\
\displaystyle{ \p_t \tilde \Psi = \left ( \frac{1}{2}f \p_x -
\frac{1}{4}\, f_{x} \right )\tilde \Psi }
\end{array}, \quad \quad
f= \p_x \log b\,, \quad f _{x} \equiv \p_x f \,, \right.
 \eeq
where we have introduced the function
$S =S(x,t)$ by the formula
 \beq\label{W1}
S={U}-\frac{{\dot b}}{2b}+\frac{b_{xx}}{2b}-\frac{3}{8} \left (\frac{b_x}{b}\right )^2=V-H -\frac{{\dot
b}}{2b}+\frac{b_{xx}}{2b}-\frac{3}{8} \left (\frac{b_x}{b}\right )^2
 \eeq
Their integrability is equivalent to the condition
 $$
\Bigl [ \frac{1}{2}\, \p_{x}^2 +S, \, \p_t -\frac{1}{2}f
\p_x +\frac{1}{4}\, f _{x}\Bigr ]\tilde \Psi =0
 $$
which implies
 \beq\label{FG2}
{\dot S} = S f_{x}+\frac{1}{2}f  S_x + \frac{1}{8}\,  f_{xxx}\,.
 \eeq
or
 \beq\label{FG203}
 \begin{array}{|c|}\hline\\
bb_{x}{U}_{x}+2bb_{x x}{U}-2b_{x}^2{U}-2{\dot{U}}b^2-\\ \
\\-\frac{1}{2}b_{x}b_{x x x}+b_{x}{\dot b}_{x}-{\dot b}^2-b{\dot
b}_{x x}+b{\ddot b}+\frac{1}{4}bb_{x x x x}+\frac{1}{4}b_{x x}^2=0\,.\\ \ \\ \hline
 \end{array}
 \eeq
This equation is of our main interest in this paper. In the next sections we determine the potential $V$ making one or
another ansatz for $b$.

Notice that the equation (\ref{FG203}) can be obtained from the
compatibility of initial matrix linear problem (\ref{zc1})
with ${U}$ defined by (\ref{potU}). One can express
all elements of ${\bf U}$ and ${\bf V}$ in terms of three functions
$a={\bf U}_{11}$, $b={\bf U}_{12}$ and ${U}$:
 \beq\label{FG204}{\bf U}:\ \ \ \ \ \ \ \ \ \ \ \
 \begin{array}{c}
{\bf U}_{11}=a\,,\ \ {\bf U}_{12}=b\,,\\ \ \\ {\bf U}_{21}=-
\frac{1}{2b^2}\left( 2a^2b+2a_xb-2ab_x+4{U}b-2{\dot b}+b_{xx}
\right)\,,
 \end{array}
 \eeq
  \beq\label{FG205}{\bf V}:\ \ \ \ \ \ \ \
 \begin{array}{c}
{\bf V}_{11}=\frac{1}{4b}\left( 2ab_x+2{\dot b}-b_{xx} \right)\,,\ \ {\bf V}_{12}=\frac{1}{2}b_x\,,\\ \ \\
{\bf V}_{21}=-\frac{1}{4b^2}\left( 4{\dot a}b-2{\dot b}_x+b_{xxx}-2ab_{xx}+2a^2b_x+4b_x{U} \right) \,.
 \end{array}
 \eeq
The function $a$ cancels out from compatibility condition (\ref{zc1}).

Recall that the dynamical variable $u$ is defined as a zero of the function $b(x,t)=b(x,u(t),t)$: $b(u,u,t)=0$. Suppose $b$ is
analytical function near $x=u$, then in the vicinity of $x=u$
 \beq\label{d21}
 \begin{array}{c}
 b=b_1(u,t)(x-u)+b_2(u,t)(x-u)^2+b_3(u,t)(x-u)^3+...
 \end{array}
 \eeq
Consider equation (\ref{FG203}) at $x=u$:
 \beq\label{FG2031}
 \begin{array}{c}
\left(-2b_{x}^2{U}-\frac{1}{2}b_{x}b_{x x x}+b_{x}{\dot b}_{x}-{\dot b}^2+\frac{1}{4}b_{x x}^2\right)\left.\right|_{x=u}=0\,,
 \end{array}
 \eeq
where we used that ${U}\!=\!V(x,t)\!-\!H({\dot u},u,t)$ and,
therefore, it is a regular function at $x=u$ and
$\left[b{U}\right](x=u)=0$. From the expansion (\ref{d21}) we get
 $$
b_x\left.\right|_{x=u}=b_1\,,\ \ b_{xxx}\left.\right|_{x=u}=6b_3\,,\ \ {\dot b}\left.\right|_{x=u}=-{\dot u}b_1\,,\ \ {\dot
b}_x\left.\right|_{x=u}={\dot u}(b_1'-2b_2)+\p_t b_1\,.
 $$
Plugging this into (\ref{FG2031}) we obtain:
 \beq\label{FG2032}
 \begin{array}{c}
{U}\left.\right|_{x=u}=-\frac{1}{2}v^2+\frac{1}{2b_1^2}\left[
\left(b_2-\frac{1}{2}b_1'\right)^2-3b_1b_3+b_1\p_t b_1
+\frac{1}{4}b_2^2 \right]\,,
 \end{array}
 \eeq
where
 \beq\label{FG2033}
v={\dot u}+\frac{b_2}{b_1}-\frac{b_1'}{2b_1}\,.
 \eeq
The latter expression is the ``momentum''.
Notice that this local evaluation at $x=u$ fixes the dependence $H({\dot u})$
since $V(x,t)$ is independent of ${\dot u}$. We consider some non-trivial  cases ($v\neq \dot u$) in Appendix A.

Let us find out what kind of restriction on the behavior of $b=b(x,u(t),t)$ arises from the Classical-Quantum Correspondence.
First, recall that the quantum Hamiltonian which we use in (\ref{scalareqs}), (\ref{W0}) has the  form $\hat
H=\frac{1}{2}\partial_x^2+V(x,t)$. Therefore, the classical one is $H(p_x,x,t)=\frac{1}{2}p_x^2+V(x,t)$. The
Classical-Quantum Correspondence implies that the classical equations for $u(t)$ arising from the compatibility condition
(\ref{P1c}) (or (\ref{FG2}) or (\ref{FG203})) are generated by
$H({\dot u}, u,t)$ which differs from $H(p_x,x,t)$ by only
possible ``quantum corrections'' of the potential. Thus, the classical Hamiltonian should have the ``Calogero form'', i. e.
$H({\dot u}, u,t)=\frac{1}{2}{\dot u}^2+{\tilde V}(u,t)$.
At the moment we do not assume any relations between $V(x,t)$ and
${\tilde V}(u,t)$. However, the Calogero form of the Hamiltonian provides some special properties of $b(x,u(t),t)$.

 \begin{predl}\label{prop0}
Let the compatibility condition (\ref{FG203}) describe
non-autonomous dynamics
 \beq\label{FG220}
  \begin{array}{c}
 {\dot u}=v\,,\\ \ \\
 {\dot v}={\ddot u}=-\p_u {\tilde V}(u,t)
  \end{array}
 \eeq
generated by the Hamiltonian
 \beq\label{FG221}
 H({\dot u}, u,t)=\frac{1}{2}{v}^2+{\tilde V}(u,t)\,.
 \eeq
Then $b(x,u(t),t)$ factorizes into the product
 \beq\label{FG222}
 b(x,u(t),t)=b_1(x-u(t),t)\, b_2(x+u(t),t)
 \eeq
and each of the factors satisfies the heat equation:
 \beq\label{FG223}
 2\p_t b_{1,2}(z,t)=\p_z^2 b_{1,2}(z,t)\,.
 \eeq
 \end{predl}
\underline{\emph{Proof}:} Substituting (\ref{FG220}) and (\ref{FG221}) into (\ref{FG203}) we get an equation where the l.h.s.
is quadratic in $v=\dot u$ . Since $v$ is an independent variable, all the coefficients in front of $v^k$ ($k=2,1,0$) vanish.
The coefficient in front of $v^2$ gives
 \beq\label{FG224}
 b_x^2-b_u^2+bb_{uu}-bb_{xx}=0
 \eeq
 or
 \beq\label{FG225}
 \left(\p_x^2-\p_u^2\right)\log b=\left(\p_x-\p_u\right)
 \left(\p_x+\p_u\right)\log b=0
 \eeq
which is equivalent to (\ref{FG222}). The coefficient
in front of $v$ gives
 \beq\label{FG226}
 b_xb_{xu}-bb_{xxu}+2bb_{tu}-2b_ub_t=0
 \eeq
or
 \beq\label{FG227}
\left(\frac{b_{xu}}{b}\right)_x=2\left(\frac{b_t}{b}\right)_u\,.
 \eeq
Plugging (\ref{FG222}) into (\ref{FG227}) we obtain:
 \beq\label{FG228}
2\left(\frac{{b_2}_t}{b_2}\right)'-2\left(\frac{{b_1}_t}{b_1}\right)'=\left(\frac{b_2''}{b_2}\right)'-
\left(\frac{b_1''}{b_1}\right)'\,.
 \eeq
The variables $x-u$ and $x+u$ are independent. Therefore,
 \beq\label{FG229}
2\left(\frac{{b_k}_t}{b_k}\right)'=\left(\frac{b_k''}{b_k}\right)'\,,\ \ k=1\,,2\,.
 \eeq
Then
 \beq\label{FG230}
2{b_k}_t={b_k''}+c(t)b_k\,,\ \ k=1\,,2\,,
 \eeq
where $c(t)$ is the integration constant. The term with $c(t)$
can be removed by the substitution $b\rightarrow b e^{\int_t
c(t)}$.\square

The coefficient in front of $v^0$ gives rise to equations for $V(x,t)$ and ${\tilde V}(u(t),t)$. We study these equations in
the next sections.

\subsection{One simple zero}

Let us first consider the case when $b$ has
only a simple zero at $u(t)$.
 The reason for this behavior of $b(z,t)$ is partly
explained in Section \ref{sec_b}.
 \begin{predl}\label{prop1}
Let $b(z,t)$ satisfy the heat equation
 \beq\label{e1}
 2\p_t b(z,t)=\p_z^2 b(z,t)\,.
 \eeq
 and let $u$ be a simple zero of the function $b$:
$b(x-u,t)\left.\right|_{x=u}=0$. Then integrability condition (\ref{FG203}) implies that
 \beq\label{e5}
 H=\frac{1}{2}\dot{u}^2+V(u)\,,
 \eeq
 \beq\label{e6}
 \ddot{u}=-V'(u)\,,
 \eeq
and
 \beq\label{e7}
  \begin{array}{|c|}
  \hline\\
 V_t(x)-V_t(u)-\frac{1}{2}f(x\!-\!u)\left(V'(x)+V'(u)\right)-
 f_x(x\!-\!u)\left(V(x)-V(u)\right)=0\,.
 \\ \ \\ \hline
 \end{array}
 \eeq
 where $f(x)=f(x,t)=b_x(x,t)/b(x,t)$ (for brevity we do not
 indicate the $t$-dependence of $f$ explicitly).
 In particular, if $f(x)=\frac{1}{x}+c_1x+c_3x^3+...$ then
 \beq\label{e17}
V'_t=\frac{1}{12}V'''+2c_1V'\,,
  \eeq
   \beq\label{e18}
\frac{1}{120}V^{(5)}=\frac{1}{2}c_1V'''+24c_3V'\,,
  \eeq
where $V_t(u)=\p_t V(u,t)$.
 \end{predl}

\underline{\emph{Proof}:} Direct substitution of $b=b(x-u(t),t)$
into (\ref{FG2}) together with (\ref{e1}) yields
 \beq\label{e2}
 V_t(x)-\dot H-\frac{1}{2}f\left(V'(x)-\ddot{u}\right)-
 f_x\left(V(x)+\frac{1}{2}\dot{u}^2-H\right)=0\,.
 \eeq
Locally,  $f=\frac{b_x}{b}\sim \frac{1}{z-u}$.
Therefore, the cancellation of the second order pole leads to (\ref{e5}). At
this stage we have
 $$
 V_t(x)-V_t(u)-\dot{u}(\ddot{u}+V'(u))-\frac{1}{2}f
 \left(V'(x)-\ddot{u}\right)-
 f_x\left(V(x)-V(u)\right)=0\,.
 $$
From the last two terms it is easy to see that the cancellation of the first order term gives (\ref{e6}). Substituting
(\ref{e6}) into the above equation we get (\ref{e7}).
The differential equations (\ref{e17}), (\ref{e18}) follows from the local expansion of (\ref{e7}) near $x=u$. To be exact,
(\ref{e18}) follows from (\ref{e17}) and $V_t'''=\frac{3}{40}V^{(5)}+\frac{5}{2}c_1V'''+24c_3V'$. $\square$

In this proof only the heat equation was used.
In what follows we need some more properties
that follow from the heat equation.
 \begin{lemma}
Let $b$ satisfy the heat equation (\ref{e1})
and $f=\p_x \log b$. Then
 \beq\label{e13}
 \p_t f=\frac{1}{2}\p_x\left (f^2+
 f_x\right )=f_x f+
 \frac{1}{2}f_{xx}\,,
 \eeq
  \beq\label{e15}
 \p_t f_x=f_{xx}f+
 f^2+\frac{1}{2}f_{xxx}\,.
 \eeq
 Suppose also that $b$ is an odd function of $x$ an has a simple zero at $x=0$. Then
   \beq\label{e16}
   \begin{array}{c}
 \frac{1}{2}f_x(x-w)f_x(x+w)=
 (f_x(x-w)+f_x(x+w))f_{xx}(2w)\\ \
 \\
 +\left ( f(x+w)-f(x-w) \right )
  f_{xx}(2w)-
 \p_tf_{x}(2w)\,.
 \end{array}
 \eeq
 \end{lemma}
\underline{\emph{Proof}:} The proof of (\ref{e13}) and (\ref{e15}) is direct.
Identity (\ref{e16}) is proved via
consideration of the local expansion and comparing of the poles
taking into account (\ref{e15}). $\square$

\subsection{Two simple zeros}

Suppose, $b$ has two simple poles. Let us derive an analogue of (\ref{e5})--(\ref{e7}) for this case.
 \begin{predl}\label{prop2}
Let $b=b_1(z,t)b_2(z,t)$ and each factor satisfies the heat equation
 \beq\label{e3}
 2\p_t b_{1,2}(z,t)=\p_z^2 b_{1,2}(z,t)\,.
 \eeq
 Suppose that $b_{1,2}$ has a simple zero $u_{1,2}$:
$b_{1,2}(x-u_{1,2},t)\left.\right|_{x=u_{1,2}}=0$. Then equation (\ref{FG2}) has the following solution:
 \beq\label{e8}
 u_1=-u_2\,,\ \ V(u)=V(-u)\,,\ \  b_1=b(x-u(t),t)\,,\ \
 b_2=b(x+u(t),t)\,,
 \eeq
 \beq\label{e10}
 H=\frac{1}{2}\dot{u}^2+V(u)+\frac{1}{2}f_x(2u)\,,
 \eeq
 \beq\label{e11}
 \ddot{u}=-V'(u)-f_{xx}(2u)\,,
 \eeq
where $b(x,t)$ is an odd function of $x$, $f=\p_x \log b$  and
the potential satisfies
 \beq\label{e9}
 \begin{array}{|c|}
   \hline\\
 V_t(x)-V_t(u)
 -\frac{1}{2}f(x\!-\!u)\left(V'(x)+V'(u)\right)
 -\frac{1}{2}f(x\!+\!u)\left(V'(x)-V'(u)\right)\\ \ \\
+\left(f_x(x\!-\!u)+f_x(x\!+\!u)\right)
 \left(V(u)-V(x)\right)=0\,.\\ \ \\ \hline
 \end{array}
 \eeq
In particular, if $f=\frac{1}{x}+c_1x+c_3x^3+...$ then
 \beq\label{e19}
V'_t=\frac{1}{12}V'''+\frac{1}{2}\, f (2x)V''+
\left(2c_1+f_x(2x)\right)V'\,,
 \eeq
  \beq\label{e20}
   \begin{array}{c}
V'''_t=\frac{3}{40}V^{(5)}+\frac{1}{2}\, f(2x)V^{(4)}+
\frac{5}{2}\left(c_1+f_x(2x)\right)V'''\\ \ \\+
\frac{9}{2}f_{xx}(2x)V''+\left(24c_3+3f_{xxx}(2x)\right)V'\,.
 \end{array}
 \eeq
 \end{predl}

\underline{\emph{Proof}:} The direct substitution leads to
 \beq\label{e4}
 \begin{array}{c}
 V_t(x)-\dot H
 -\frac{1}{2}\frac{b_{1,x}}{b_1}\left(V'(x)-\ddot{u}_1\right)-\frac{1}{2}\frac{b_{2,x}}{b_2}\left(V'(x)-\ddot{u}_2\right)\\
 \ \\-\left(\frac{b_{1,x}}{b_1}\right)_x\left(V(x)+\frac{1}{2}\dot{u}_1^2-H+\frac{1}{2}(\dot{u}_1+\dot{u}_2)\frac{b_{2,x}}{b_2}\right)
 \\
 \ \\-\left(\frac{b_{2,x}}{b_2}\right)_x\left(V(x)+\frac{1}{2}\dot{u}_2^2-H+\frac{1}{2}(\dot{u}_1+\dot{u}_2)\frac{b_{1,x}}{b_1}\right)
 \\ \ \\-\frac{1}{2}\left(\frac{b_{1,x}}{b_1}\right)_x\left(\frac{b_{2,x}}{b_2}\right)_x=0\,.
 \end{array}
 \eeq
From cancellation of the second order poles we get
 \beq\label{e12}
  \begin{array}{c}
H=\frac{1}{2}\dot{u}_1^2+V(u_1)+\frac{1}{2}(\dot{u}_1+\dot{u}_2)\frac{b_{2,x}}{b_2}(u_1)
+\frac{1}{2}\left(\frac{b_{2,x}}{b_2}\right)_x(u_1)\,,\\ \ \\
H=\frac{1}{2}\dot{u}_2^2+V(u_2)+\frac{1}{2}(\dot{u}_1+\dot{u}_2)\frac{b_{1,x}}{b_1}(u_2)
+\frac{1}{2}\left(\frac{b_{1,x}}{b_1}\right)_x(u_2)\,.
 \end{array}
 \eeq
Comparing these two expressions one can see that (\ref{e8}) and (\ref{e10}) indeed satisfy (\ref{FG2}). Then vanishing of the
first order poles at $\pm u$ gives (\ref{e11}).
Substituting (\ref{e11}) into (\ref{e4}) we get
 \beq\label{e14}
 \begin{array}{c}
 V_t(x)-\left. \p_t\left(V(z)+\frac{1}{2}f(2z)\right)\right|_{z=u(t)}
 \\  \\-\frac{1}{2}\frac{b_{1,x}}{b_1}\left(V'(x)+V'(u)+f_{xx}(2u)\right)
 -\frac{1}{2}\frac{b_{2,x}}{b_2}\left(V'(x)-V'(u)-f_{xx}(2u)\right)\\
 \ \\+\left(\left(\frac{b_{1,x}}{b_1}\right)_x+
 \left(\frac{b_{2,x}}{b_2}\right)_x\right)
 \left(V(u)-V(x)+\frac{1}{2}f_{xx}(2u)\right)
-\frac{1}{2}\left(\frac{b_{1,x}}{b_1}\right)_x
\left(\frac{b_{2,x}}
{b_2}\right)_x=0\,.
 \end{array}
 \eeq
All terms that do not contain $V$ cancel because of (\ref{e16})
and we get (\ref{e9}). Differential equations
(\ref{e19}), (\ref{e20}) follows from the local expansion of (\ref{e9}) near $x=u$. $\square$

\noindent
{\bf Remark.} To investigate the case more general than
(\ref{e8})) one should solve the equation emerging from equality of right
hand sides of (\ref{e12}) (see Appendix A).

Notice also that the r.h.s. of (\ref{e19}) and
(\ref{e20}) are full derivatives:
 \beq\label{e191}
V'_t=\p_x\left(\frac{1}{12}V''+2c_1V+ \frac{1}{2}
f(2x)V'\right)\,,
 \eeq
  \beq\label{e201}
   \begin{array}{c}
V'''_t=\p_x\left(\frac{3}{40}V^{(4)}+\frac{5}{2}c_1V''+24c_3V \right.\\ \ \\+
\left.\frac{3}{2}f_{xx}(2x)V'+\frac{3}{2}
f_{x}(2x)V''+
\frac{1}{2}f(2x)V'''\right)\,.
 \end{array}
 \eeq
In particular, this leads to the following equation:
  \beq\label{e203}
   \begin{array}{c}
V^{(4)}-60c_1V''+60f_{xx}(2x)V'+60f_{x}(2x)V''+24c_3V=\hbox{const}(t)\,.
 \end{array}
 \eeq

\section{Rational Solutions}
\setcounter{equation}{0}

\subsection{The simplest case: $b=x-u(t)$}

The simplest possibility is to set
 \beq\label{S1}
b=x-u(t)\,.
 \eeq
We will see that already this case is meaningful and leads
to ${\rm P}_{\rm I}$ and ${\rm P}_{\rm II}$ equations.

In this case  integrability condition (\ref{e7}) turns into
 \beq\label{S403}
\Bigl (V_t(x)-V_t(u)\Bigr )-\frac{1}{2(x-u)}
\Bigl (V'(x)+V'(u)\Bigr )+\frac{1}{(x-u)^2}\Bigl (V(x)-V(u)\Bigr )=0
 \eeq
or
 \beq\label{S5} 2(x-u)^2 \Bigl (V_t(x)-V_t(u)\Bigr )-(x-u)\Bigl (V'(x)+V'(u)\Bigr )+ 2\Bigl (V(x)-V(u)\Bigr )=0
 \eeq
It should be an identity for all $x,u$ which enter here as
independent variables on equal footing.
The way to proceed is to take the third derivative of
(\ref{S5}) with respect to $x$.
The result is
 $$
\begin{array}{c}
2u^2 V_{t}'''(x) +u \left (V^{\rm IV}(x)-4x V_{t}'''(x)
-12V_{t}''(x)\right )
\\ \\
\displaystyle{
+\,\,\,\,
12 V_{t}'(x)+12x V_{t}''(x)+2x^2 V_{t}'''(x)-V'''(x)-x V^{\rm IV}(x)
=0}
\end{array}
 $$
The equality holds identically if
the coefficients in front of $u^2$, $u$ and the free term in $u$ vanish.
This implies the conditions
 \beq\label{S6}
\left \{
\begin{array}{l}
V_{t}'''(x)=0 \\ \\
12V_{t}'(x) =V'''(x)
\end{array}\right.
 \eeq
From the first equation it follows that $V_{t}(x)$
is a polynomial in $x$ of second degree at most while
from the second one it then follows that $V(x)$ is
a polynomial in $x$ of fourth degree at most.
There are three possibilities:
\begin{itemize}
\item[1)] $V_{t}'(x)\equiv 0$, then $V(x)$ is a quadratic
polynomial $V(x)=a_2 x^2 +a_1 x +a_0$ with $\dot a_2 =\dot a_1 =0$.
Plugging it into equation (\ref{S5}), we see that the equation
holds identically for any constants $a_2, a_1$, with the irrelevant
free term $a_0$ being
an arbitrary function of $t$. This is the potential for the
harmonic oscillator.
\item[2)] $V_{t}''(x)\equiv 0$, then $V(x)$ is a 3-d degree polynomial
$V(x)=a_3 x^3 +a_2 x^2 +a_1 x +a_0$ with $\dot a_3 =\dot a_2 =0$. By rescaling and shift of the variable $x$ we can set
$a_3=1$, $a_2=0$. The free term, $a_0$, is irrelevant since it cancels in equation (\ref{S5}). Plugging the potential in the
form $V(x)=x^3 +a_1 x$ into equation (\ref{S5}), we get $(x-u)^2 (2\dot a_1 -1)=0$. Therefore, $a_1 =t/2$ and
 $$
V(x)=x^3 +\frac{tx}{2}
 $$
This is, up to a common factor, the potential for the
${\rm P}_{\rm I}$ equation.
\item[3)] $V_{t}''(x)\neq 0$, then $V(x)$ is a 4-th degree polynomial
$V(x)=a_4 x^4 +a_3 x^3 +a_2 x^2 +a_1 x +a_0$ with
$\dot a_4 =\dot a_3  =0$.
Again, we can set $a_4=1$, $a_3=0$ and $a_0 =0$.
Plugging the potential in the form
$V(x)=x^4 +a_2 x^2 +a_1 x$ into equation (\ref{S5}), we get
$(x^2 -u^2)(\dot a_2 -1) +\dot a_1 =0$. Therefore, $a_2 =t$,
$a_1 =-2\alpha$, where $\alpha$ is an arbitrary constant.
Up to a common factor, we obtain the potential
 $$
V(x)=x^4 +tx^2 -2\alpha x
 $$
for the
${\rm P}_{\rm II}$ equation with the parameter $\alpha$.
\end{itemize}

\subsection{The case $b=(x-u_1(t))(x-u_2(t))$}

Let us make the similar calculations for $b=(x-u_1 (t))(x-u_2(t))$. Instead of
 $$ V_t(x)-H_t -\frac{V'(x)-\ddot
u}{2(x-u)} +\frac{V(x)-H +\dot u^2/2}{(x-u)^2}=0
 $$
 we get, after cancellation of third and fourth order poles:
 \beq\label{S7}
\begin{array}{c}
V_t(x)-H_t -\frac{1}{2(x-u_1)}\Big(V'(x)-\ddot
u_1-\frac{2}{(u_1-u_2)^3}\Big) -\frac{1}{2(x-u_2)}\Big(V'(x)-\ddot
u_2-\frac{2}{(u_2-u_1)^3}\Big)\\ \\ +\frac{1}{(x-u_1)^2}\Big( V(x)-H
+\frac{1}{2}\frac{\dot u_1+\dot u_2}{u_1-u_2}+\frac{1}{2}\dot
u_1^2-\frac{1}{2(u_1-u_2)^{2}}\Big)\\ \\ \ \ \ \ \
+\frac{1}{(x-u_2)^2}\Big( V(x)-H +\frac{1}{2}\frac{\dot u_1+\dot
u_2}{u_2-u_1}+\frac{1}{2}\dot u_2^2-\frac{1}{2(u_1-u_2)^{2}}\Big)=0
\end{array}
 \eeq
Cancellation of the second order poles at $x=u_{1,2}$ yields
 \beq\label{S703} H=\frac{1}{2}\dot u_1^2+\frac{1}{2}\frac{\dot
u_1+\dot u_2}{u_1-u_2}+V(u_1)-\frac{1}{2(u_1-u_2)^{2}} \eeq and
\beq\label{S704} H=\frac{1}{2}\dot u_2^2 +\frac{1}{2}\frac{\dot
u_1+\dot u_2}{u_2-u_1}+V(u_2)-\frac{1}{2(u_1-u_2)^{2}}
 \eeq
By equating the two ``kinetic'' terms we get the following
two possibilities:
 \beq\label{S705}
{\bf\hbox{1)}}\ \ \dot u_1+\dot u_2=0 ,
\hspace{33mm}
{\bf\hbox{2)}}\ \ \p_t(u_1-u_2)^2=-4
 \eeq
In the first case $u_1+u_2=\hbox{const}$ and one can shift $x$ in
the initial problem to set $u_1=-u_2\equiv u$. Therefore, the
two possibilities are rewritten as
 \beq\label{S706}
\hspace{27mm} {\bf\hbox{1)}}\ \
\left\{\begin{array}{l}u_1=-u_2\equiv u\\ \\ V(u)=V(-u)
\end{array}\right. \hspace{18mm} {\bf\hbox{2)}}
\ \ \left\{\begin{array}{l}u_1=u_2+\sqrt{c-4t} \\ \\
V(u)=V(u-\sqrt{c-4t}),
\end{array}\right.
 \eeq where $c$ is some constant. The second case is given in the Appendix A. Here we consider the first one.
In this case (\ref{S7}) leads to integrability condition (\ref{e9}):
 \beq\label{S10}
\begin{array}{c}
\displaystyle{
V_t(x)-V_t(u)-\frac{1}{2(x-u)}\left (V'(x)+V'(u) -2\,
\frac{V(x)-V(u)}{x-u}\right )}
\\ \\
\displaystyle{
-\,\,\,
\frac{1}{2(x+u)}\left (V'(x)-V'(u) -2\, \frac{V(x)-V(u)}{x+u}
\right )=0}
\end{array}
 \eeq
or, equivalently,
 \beq\label{S10a}
\begin{array}{c}
\displaystyle{2(x^2-u^2)^2(V_t(x)-V_t(u))- (x+u)(x^2-u^2)\left (V'(x)+V'(u)\right )}
\\ \\
\displaystyle{-\,\,
(x-u)(x^2-u^2)\left (V'(x)-V'(u)\right )
+4(x^2+u^2)(V(x)-V(u))\, =\, 0}
\end{array}
 \eeq

Since the maximal degree of $x$ in (\ref{S10a}) is $4$,
the differential operator $\p_{X}^{5}$ applied to this equation
kills all terms containing $V(u)$ and we are left with
 $$
\p_{X}^{5}\Bigl [(x^2-u^2)^2 V_t (x)-x(x^2 -u^2)
V'(x)+2(x^2 +u^2)V(x)\Bigr ]=0
 $$
Equating the coefficients in front of $u^4$, $u^2$ and $u^0$ to zero,
we get the following conditions:
 $$
\left \{\begin{array}{l}
\displaystyle{
\p_{x}^{5} V_t (x)=0}
\\ \\
\displaystyle{
\p_{x}^{5}\Bigl [-2x^2 V_t (x)+x V'(x) +2V(x) \Bigr ]=0}
\\ \\
\displaystyle{
\p_{x}^{5}\Bigl [ x^4 V_t (x)-x^3 V'(x)
+2x^2 V(x)  \Bigr ]=0}
\end{array}\right.
 $$
They mean that the expressions in the square brackets are
polynomials in $x$ of at most fourth degree:
 $$
\begin{array}{rll}
V_t (x)&=&P_4(x)
\\ \\
-2x^2 V_t (x)+x V'(x) +2V(x) &=&Q_4(x)
\\ \\
x^4 V_t (x)-x^3 V'(x)
+2x^2 V(x)  &=& R_4(x)
\end{array}
 $$
Combining these conditions, we find that
$x^2 V(x)$ must be a polynomial
of at most $8$-th degree such that its highest
and lowest coefficients do not depend on $t$.
We also recall that it must contain
only even powers of $x$. So we can write
 $$
V(x)=\mu x^6 +a_4x^4 +a_2x^2 +a_0 +\frac{\nu}{x^2}\,, \quad \quad
\dot \mu =\dot \nu =0.
 $$
Plugging this potential back to equation (\ref{S10a}),
we obtain
 $$
(x^4-u^4)(x^2 -u^2)(\dot a_4-4\mu )+(x^2-u^2)^2(\dot a_2-2a_4)=0
 $$
The solution is $a_4 =4\mu t +\alpha_4 $,
$a_2 = 4\mu t^2 +2\alpha _4 t+\alpha_2$
with integration constants $\alpha_4$, $\alpha_2$ and $a_0$
is arbitrary. There are three cases:
 \begin{itemize}
\item[1)] $\mu \neq 0$ (the case of
general position), then one can put it equal to $1$ by rescaling
and set $\alpha_4=0$ by a shift of the $t$-variable.
Then the potential acquires the form
 \beq\label{S11}
V(x,t)=x^6 +4tx^4 +(4t^2 +\alpha_2)x^2 +a_0(t) +
\frac{\nu}{x^2}
 \eeq
This is the potential for the ${\rm P}_{\rm IV}$
equation.
\item[2)] $\mu = 0$ but $\alpha_4 \neq 0$,
then one can put $\alpha_4$
equal to $1$ by rescaling and set
$\alpha_2=0$ by a shift of the $t$-variable.
The potential is
 \beq\label{S11a}
V(x,t)=x^4 +2tx^2 +a_0(t) + \frac{\nu}{x^2}
 \eeq
It generates the equation
 \beq\label{S11ab}
\ddot u =-4u^3 -4tu +\frac{2\nu}{u^3}
 \eeq
The change of the dependent variable
$u\to y$ such that $u^2+
y^2+\frac{1}{2}\dot y+t=0$
(a version of
a similar change in \cite[section 14.331]{Ince}) brings the equation to
the form $\ddot y =8y^3 +8ty +\sqrt{-32\nu}-2$
which is equivalent to
the ${\rm P}_{\rm II}$ equation.
\item[3)] $\mu =\alpha_4 = 0$, then
 \beq\label{S11b}
 V(x,t)=\alpha_2 x^2+\frac{\nu}{x^2} \, +a_0(t)
 \eeq
 This gives the exactly solvable rational 2-particle
Calogero model in the harmonic potential. The $x$-independent term $a_0(t)$ is irrelevant.
 \end{itemize}

\section{Hyperbolic Solutions}
\setcounter{equation}{0}

\subsection{The case $b=e^{t/2} \sinh (x-u(t))$}

Let us consider the case when $b$ is a trigonometric (hyperbolic, to be exact) function with one simple zero in the strip of
periodicity:
 \beq\label{SS1} b=e^{t/2} \sinh (x-u(t))\,.
  \eeq
 We will see that it leads to the ${\rm P}_{\rm III}$ equation. Since $b$ satisfies the heat equation (\ref{e1}), the
 Proposition \ref{prop1} can be applied. The
integrability condition (\ref{e7}) with $\frac{b_x}{b}=\coth(x)$ becomes:
 \beq\label{SS5}
 \begin{array}{c}
 2\sinh ^2 (x-u) \Bigl (V_t(x)-V_t(u)\Bigr )-\sinh (x-u)\cosh (x-u) \Bigl
(V'(x)+V'(u)\Bigr )\\ \\ +2\Bigl (V(x)-V(u)\Bigr )=0\,.
 \end{array}
 \eeq

Let us make the change of variables $V \to {\cal V}$,
$x \to X$, $u \to U$ such that
 $$
V(x)\equiv e^{-4x}{\cal V}(e^{2x})\,, \quad X=e^{2x}\,,
\quad U=e^{2u}
 $$
then equation (\ref{SS5}) is rewritten as
 \beq\label{SS5a}
\begin{array}{c}
(X-U)^2 \left (U^2 {\cal V}_t (X)-X^2 {\cal V}_t (U)\right )
-UX(X^2 -U^2)\left ( U{\cal V}'(X)+ X{\cal V}'(U)\right )
\\ \\
+\,\, 2(X^2 -U^2)\left ( U^2 {\cal V}(X)+X^2 {\cal V}(U)\right ) +4UX \left ( U^2 {\cal V}(X)-X^2 {\cal V}(U)\right )\, =\,
0\,.
\end{array}
 \eeq
Since the maximal degree of $X$ here equals  $4$, the differential operator $\p_{X}^{5}$ applied to this equation kills all
terms containing ${\cal V}(U)$ and we are left with
 $$
\p_{X}^{5}\Bigl [(X-U)^2 {\cal V}_t (X)-X(X^2 -U^2) {\cal V}'(X)+2(X^2 -U^2+2UX){\cal V}(X)\Bigr ]=0\,.
 $$
Equating the coefficients in front of $U^2$, $U^1$ and $U^0$ to zero,
we get the following conditions:
 \beq\label{SS6}
\left \{\begin{array}{l} \displaystyle{ \p_{X}^{5}\Bigl [{\cal V}_t (X)+X{\cal V}'(X)-2{\cal V}(X)\Bigr ]=0}\,,
\\ \\
\displaystyle{ \p_{X}^{5}\Bigl [-X  {\cal V}_t (X)+2X {\cal V}(X)  \Bigr ]=0}\,,
\\ \\
\displaystyle{ \p_{X}^{5}\Bigl [ X^2 {\cal V}_t (X)-X^3 {\cal V}'(X) +2X^2{\cal V}(X)  \Bigr ]=0}\,.
\end{array}\right.
 \eeq
They mean that the expressions in the square brackets are
polynomials in $X$ of at most fourth degree:
 \beq\label{SS6a}
\begin{array}{l}
\displaystyle{ {\cal V}_t (X)+X{\cal V}'(X)-2{\cal V}(X)=P_4(X)}\,,
\\ \\
\displaystyle{ -X  {\cal V}_t (X)+2X {\cal V}(X)  =Q_4(X)}\,,
\\ \\
\displaystyle{ X^2 {\cal V}_t (X)-X^3 {\cal V}'(X) +2X^2{\cal V}(X) = R_4(X)}\,.
\end{array}
 \eeq
Combining these conditions, we obtain that $X^2 {\cal V}'(X)$ and $X^2 {\cal V}_t(X)$ are polynomials of at most $5$-th and
$6$-th degrees respectively. It is easy to see that the former polynomial must be divisible by $X^2$. Indeed, let it be $X^2
{\cal V}'(X)=X^2 P_3(X) +p_1 X +p_0$ with some nonzero $p_{0,1}$, then the first equation in (\ref{SS6a}) implies $p_0=0$
(otherwise the left hand side contains a non-polynomial term $\propto X^{-1}$) and the second equation multiplied by $X$
implies $p_1 =0$ (otherwise the left hand side contains a non-polynomial term $\propto X^{2}\log X$). Therefore, we conclude
that ${\cal V}'(X)$ is a polynomial of at most third degree and, thus, ${\cal V}(X)$ itself is a polynomial of at most fourth
degree:
 $$
{\cal V}(X)=a_4 X^4 +a_3 X^3 +a_2 X^2 +a_1 X +a_0\,.
 $$
Let us plug it in equation (\ref{SS5a}). After simple transformations we obtain the relation:
 $$
\begin{array}{c}
(X-U)(X^2-U^2)(\dot a_4 -2a_4)+(X-U)^2 (\dot a_3 -a_3)
\\ \\
\displaystyle{ -\,\,\, \frac{(X-U)^2}{UX}\, (\dot a_1 -a_1)- \frac{(X-U)(X^2-U^2)}{U^2X^2}\, (\dot a_0 -2a_0)\, =\, 0}\,.
\end{array}
 $$
It must be satisfied identically for all $X,U$. This implies $\dot a_4 =2a_4$,  $\dot a_3 =a_3$, $\dot a_1 =a_1$, $\dot a_0
=2a_0$ and no condition for $a_2$. Therefore, the potential $V(x,t)$ is fixed to be
 \beq\label{SS7}
V(x,t)=\alpha _1 e^{2t +4x} +\alpha_2 e^{2t -4x} +\alpha _3 e^{t +2x} +\alpha_4 e^{t -2x} +a(t)\,,
 \eeq
where $\alpha_i$ are arbitrary constants. This is precisely
the potential for the ${\rm P}_{\rm III}$ equation.

\subsection{The case $b=e^{t}\sinh(x-u(t))\sinh(x+u(t))$}
In this case $b=\left(e^{t/2}\sinh(x-u)\right)\left(e^{t/2}\sinh(x+u)\right)$. Each of the multiples satisfies the heat
equation (\ref{e3}). Therefore the Proposition \ref{prop2} can be applied. Then equation (\ref{e9}) assumes the form
 \beq\label{S23}
 \begin{array}{c}
 V_t(x)-V_t(u) -\frac{1}{2}\coth(x-u)\Big( V'(x)+V'(u)\Big) -\frac{1}{2}\coth(x+u)\Big(
V'(x)-V'(u) \Big)
\\ \\
+\Big(V(x)-V(u)\Big)\Big(\frac{1}{\sinh^2(x-u)}+\frac{1}{\sinh^2(x+u)}\Big)=0\,.
\end{array}
 \eeq
Multiplying by $32\sinh^2(x-u)\sinh^2(x+u)$ and making change of variables $X=\cosh^2(x)$, $y=\coth^2(u)$ we get
 \beq\label{S24}
 \begin{array}{c}
 (Xy-X-y)^2(V_t(X)-V_t(y)) -2X(X-1)(y-1)(Xy-X-y)V'(X)+\\ \ \\+2y(y-1)(Xy-X-y)V'(y)+2(y-1)(Xy+X-y)(V(X)-V(y))=0\,.
\end{array}
 \eeq
Now one can apply the calculation method similar to the previous cases. That is to take the third derivative with respect to
$X$ and analyze the differential equations (the later equations appear as the coefficients behind different powers of $y$).
This analysis gives the potential of the Painlev\'e V equation after some tedious evaluations. Instead of doing in this
manner, let us simplify the problem by assuming that the solution is a sum of terms of the form $V(x)=e^{kt}v(X)$. Making
this substitution one gets:
 \beq\label{S26}
 \begin{array}{c}
 k(Xy-X-y)^2(v(X)-v(y)) -2X(X-1)(y-1)(Xy-X-y)v'(X)+\\ \ \\+2y(y-1)(Xy-X-y)v'(y)+2(y-1)(Xy+X-y)(v(X)-v(y))=0\,.
\end{array}
 \eeq
We will see that nontrivial solutions exist for $k=0,2,4$.
The way to proceed is to take the third derivative of the expression with respect to $X$. The equality holds identically if
the coefficients in front of $y^2$, $y$ and the free term in $y$ vanish. This implies the following conditions:
 \beq\label{S25}
 \left\{
 \begin{array}{l}
 X(X-1)v'''(X)+3(2X-1)v''(X)+3(2-k)v'(X)=0\,,\\ \ \\
 X(X-1)v^{(4)}(X)+4(2X-1)v'''(X)+3(4-k)v''(X)=0\,,\\ \ \\
 kv'''(X)=0\,.
 \end{array}
 \right.
 \eeq
Consider  the last equation. If $k=0$ one gets $$v(X)=\frac{c_1}{X}+\frac{c_2}{X-1}+c_3=\frac{\tilde{c}_1}{\sinh^2
x}+\frac{\tilde{c}_2}{\cosh^2 x}+\tilde{c}_3\,,$$ else $v'''(X)=0$. The later case leads to $(k-4)v''(X)=0$ (from the second
equation in (\ref{S25})). Then, $k=4$ or $v''(X)=0$. In the later case one gets $(k-2)v'(X)=0$ (from the first equation in
(\ref{P5aaab})).
In this way one can easily recover the potential of the Painlev\'e V equation (\ref{P5aaab}):
 \beq\label{P5e4}
\begin{array}{c}
V(x,t)=-\frac{2(\xi +\sigma )^2}{\sinh^2 x} +\frac{2\zeta^2}{\cosh^2 x}+ \frac{e^{2t}}{2}(2\sigma -1)\cosh (2x )-
\frac{e^{4t}}{16}\cosh (4x  )\,.
\end{array}
\eeq

\section{Elliptic Solutions}
\setcounter{equation}{0}

\subsection{The case $b=\vth(x-u(t),2\pi i t)$}
Consider an elliptic curve  with moduli $\tau=2\pi  i t$ $$\Sigma_\tau:\ \ {\CC}/{\ZZ}+{\ZZ}\tau$$ and let
$b=\vth(x-u(t),2\pi i t)$. Definitions and properties of elliptic functions are given in the Appendix B. Then from (\ref{e7})
we have
 \beq\label{S37} \begin{array}{c} V_t(x)-V_t(u)-\frac{1}{2}E_1(x-u)(V'(u)+V'(x))
 +E_2(x-u)\left(V(x)-V(u)\right)=0\,.
 \end{array}\eeq
We will show that this equation has only trivial solutions $V(x,t)=f(t)$. For this purpose consider the same equation at
$x+\tau$ and subtract it from the initial one. Then, using the behavior of $E_1(z)$ (\ref{A.102}) and $E_2(z)$ (\ref{A.103})
on the torus lattice we get:
 \beq\label{S3721}
 \begin{array}{c}
 V_t(x+\tau)-V_t(x)-\frac{1}{2}E_1(x-u)(V'(x+\tau)-V'(x))\\ \\ +E_2(x-u)(V(x+\tau)-V(x))
 +\pi i (V'(u)+V'(x+\tau))=0\,.
 \end{array}
 \eeq
Let us now differentiate the obtained equality with respect to $x$:
 \beq\label{S3722}
 \begin{array}{c}
 V'_t(x+\tau)-V'_t(x)-\frac{1}{2}E_1(x-u)(V''(x+\tau)-V''(x))+\pi i V''(x+\tau)\\ \\ +\frac{3}{2}E_2(x-u)(V'(x+\tau)-V'(x))
 +E'_2(x-u)(V(x+\tau)-V(x))=0\,.
 \end{array}
 \eeq
Similarly, let us shift the argument $u\rightarrow u+\tau$ in the equation (\ref{S3722}) and subtract it from (\ref{S3722})
itself (keeping in mind that $E_2'$ is the double-periodic function). This gives
 \beq\label{S3723}
 \begin{array}{c}
V''(x+\tau)-V''(x)=0
 \end{array}
 \eeq
or
 \beq\label{S3724}
 \begin{array}{c}
V(x+\tau)-V(x)=a(\tau)x+b(\tau)\,.
 \end{array}
 \eeq
Plugging this back to (\ref{S3721}) one can easily get that $a(\tau)=b(\tau)=0$ by analyzing coefficients behind the poles at
$x-u$ of the second and the first orders. Therefore, the potential should be a double-periodic function. If it is, then
(\ref{S3721}) reduces to
 $$
 \begin{array}{c}
 \p_t(V(x+\tau)-V(x))
 +\pi i (V'(u)-V'(x))=0\,.
 \end{array}
 $$
since $\p_t(V(x+\tau)-V(x))=V_t(x+\tau)-V_t(x)+2\pi i V'(x+\tau)$. The later equation should hold for all $x$ and $u$. Then
the only solution is $$V(x,t)=f(t)\,.$$

\subsection{The case $b=\vth(x-u(t),2\pi i t)\vth(x+u(t),2\pi i t)$}

Equation (\ref{e9}) in this case has the form
 \beq\label{S55} \begin{array}{c} V_t(x)-V_t(u)
 -\frac{1}{2}E_1(x\!-\!u)(V'(u)+V'(x))-\frac{1}{2}E_1(x\!+\!u)(-V'(u)+V'(x))
 \\ \ \\
+(E_2(x\!-\!u)+E_2(x\!+\!u))(V(x)-V(u))=0\,.
 \end{array}\eeq
Let us make a change of variables:
 \beq\label{s60}
X(x,t)=\frac{\wp(x)-e_1}{e_2-e_1}\,,\ \  Q(u,t)=\frac{\wp(u)-e_1}{e_2-e_1}\,,\ \ T(t)=\frac{e_3-e_1}{e_2-e_1}\,.
 \eeq
Then
 \beq\label{s61}
E_1(x+u)+E_1(x-u)=2E_1(x)+\frac{\wp'(x)}{\wp(x)-\wp(u)}=2E_1(x)+\frac{X_x}{X-Q}\,,
 \eeq
  \beq\label{s62}
E_1(x+u)-E_1(x-u)=2E_1(u)+\frac{\wp'(u)}{\wp(u)-\wp(x)}=2E_1(x)-\frac{Q_u}{X-Q}\,,
 \eeq
 \beq\label{s63}
E_2(x+u)+E_2(x-u)=2E_2(u)+\frac{Q_{uu}}{X-Q}+\frac{Q_u^2}{(X-Q)^2}\,.
 \eeq
Therefore, equation (\ref{S55}) is written as
 \beq\label{s64} \begin{array}{c} (V_T(X)-V_T(Q))T_t+V_X(X)X_t-V_Q(Q)Q_t
\\ \ \\ -\frac{1}{2}V_X(X)X_x\left(2E_1(x)+\frac{X_x}{X-Q}\right)
+\frac{1}{2}V_Q(Q)Q_u\left(2E_1(u)-\frac{Q_u}{X-Q}\right) \\ \ \\ +\left(2\left[2\eta_1+e_1+(e_2-e_1)Q\right]+
\frac{Q_{uu}}{X-Q}+\frac{Q_u^2}{(X-Q)^2}\right)(V(x)-V(u))=0\,.
 \end{array}\eeq
It follows from  (\ref{Bp27}) that
 $$X_t-X_xE_1(x)=X_x(E_1(x+\om_3)\!-\!E_1(x)\!-\!E_1(\om_3))=\frac{1}{2}X_x\frac{\wp'(x)}{\wp(x)-\wp(\om_3)}=
\frac{1}{2}\frac{X_x^2}{X-T}\,.
 $$
Therefore,
 \beq\label{s65} \begin{array}{c} (V_T(X)-V_T(Q))T_t+V_X(X)\frac{1}{2}\frac{X_x^2}{X-T}-V_Q(Q)\frac{1}{2}\frac{Q_u^2}{Q-T}
\\ \ \\ -\frac{1}{2}V_X(X)\frac{X_x^2}{X-Q}
-\frac{1}{2}V_Q(Q)\frac{Q_u^2}{X-Q} \\ \ \\ +\left(2\left[2\eta_1+e_1+(e_2-e_1)Q\right]+
\frac{Q_{uu}}{X-Q}+\frac{Q_u^2}{(X-Q)^2}\right)(V(x)-V(u))=0\,.
 \end{array}\eeq
Now let us proceed as in the previous examples. First, multiply (\ref{s65}) by $(X-Q)^2$. Secondly, take the third derivative
with respect to $X$. This excludes $V(Q)$. Thirdly, substitute $Q_u^2=4(e_2-e_1)Q(Q-1)(Q-T)$ and
$Q_{uu}=2(e_2-e_1)(3Q^2-2Q(T+1)+T)$. Then, the coefficients in front of $Q^2$, $Q^1$ and $Q^0$ should vanish independently:
 \beq\label{s66}
 \left\{\begin{array}{l}
 F+2V(X)(2\eta_1+e_1+X(e_2-e_1))=P_2(X)\,,\\ \ \\
 -2XF+\frac{1}{2}V_X(X)X_x^2\\ \ \\+2V(X)(X^2(e_2-e_1)+4X(e_1-\eta_1)+T(e_2-e_1))=Q_2(X)\,,\\ \ \\
 X^2 F-\frac{1}{2}V_X(X)X_x^2X+2V(X)((2\eta_1+e_1)X^2+(e_2-e_1)XT)=R_2(X)\,,
 \end{array}\right.\eeq
where $P_2(X)$, $Q_2(X)$, $R_2(X)$ are the second order polynomials in $X$ with times-dependent coefficients and
$F=V_T(X)T_t+V_X(X)\frac{1}{2}\frac{X_x^2}{X-T}$.

Excluding $F$ from two upper equations in (\ref{s66}) we obtain the following equality:
 \beq\label{s67} \begin{array}{c} V_X(X)X(X\!-\!1)(X\!-\!T)+V(X)\Big(X(X\!-\!1)+X(X\!-\!T)+(X\!-\!1)(X\!-\!T)\Big)\\ \ \\
 =\frac{1}{e_2-e_1}\Big(\frac{1}{2}Q_2(X)+XP_2(X)\Big)\,.
 \end{array}\eeq
General solution of the later equation has a form:
 \beq\label{s68} \begin{array}{c} V(X)=\frac{1}{X(X-1)(X-T)}\int^XdZ\frac{1}{e_2-e_1}\Big(\frac{1}{2}Q_2(Z)+ZP_2(Z)\Big)=
 \frac{H_4(X)}{X(X-1)(X-T)}\,,
 \end{array}\eeq
where $H_4(X)$ is the forth order polynomials in $X$ with times-dependent coefficients. Therefore, $V(X)$ can be presented as
 \beq\label{s69} \begin{array}{c} V(X)=a(T)X+\frac{b(T)}{X}+\frac{c(T)}{X-1}+\frac{d(T)}{X-T}+h(T)\,.
 \end{array}\eeq
The last term $h(T)$ is not fixed by (\ref{S55}), i. e. $h(T)$ is arbitrary.

Plugging (\ref{s69}) into (\ref{s65}) and multiplying the result by $(X-Q)X(X-1)(X-T)Q(Q-1)(Q-T)$ we get a polynomial
function in $X$ and $Q$. The coefficients in front of $Q^kX^j$ provides differential equations. It can be verified that all
of them are equivalent to the following system:
 \beq\label{s70} \left\{\begin{array}{l}
 a_T(T) T(T-1)(e_2-e_1)+a(T)(e_3+2\eta_1)=0\,,\\ \ \\
 b_T(T) T(T-1)(e_2-e_1)+b(T)(e_2+2\eta_1)=0\,,\\ \ \\
 c_T(T) T(T-1)(e_2-e_1)+c(T)(e_1+2\eta_1)=0\,,\\ \ \\
 d_T(T) T(T-1)(e_2-e_1)+d(T)(-2e_3+2\eta_1)=0\,.
 \end{array}\right.
 \eeq
Its solutions (see (\ref{6p8})-(\ref{6p88})) are
 \beq\label{s71} \left\{\begin{array}{l}
 a(T)=\alpha(e_2-e_1)\,, \ \ \alpha=\hbox{const}\,,\\ \ \\
 b(T)=\beta(e_2-e_1)T\,, \ \ \beta=\hbox{const}\,,\\ \ \\
 c(T)=\gamma(e_2-e_1)(T-1)\,, \ \ \gamma=\hbox{const}\,,\\ \ \\
 d(T)=\delta(e_2-e_1)T(T-1)\,, \ \ \delta=\hbox{const}\,.
 \end{array}\right.
 \eeq
Then, in view of (\ref{6p7}) we have
 \beq\label{s72} \begin{array}{c} V(x)=\alpha\wp(x)+\beta\wp(x+\om_1)+\gamma\wp(x+\om_2)+\delta\wp(x+\om_3)+h(t)\,.
 \end{array}\eeq

This is the potential of the Painlev{\'e} VI equation in the elliptic form \cite{Painleve1906, Inoz, Manin98,Zotov04} (see
also \cite{LOZ2,Takasaki02} and \cite{Dubrovin2}). We remark that the non-stationary Lam\'e equation in connection with the
${\rm P}_{\rm VI}$ equation (and with the 8-vertex model) was discussed in \cite{BM06}. Recently, the non-stationary Lam\'e
equation has appeared \cite{FL,MMM}, \cite{agt1,agt2,agt3} in the context of the AGT conjecture. The results of
\cite{agt1,agt2,agt3} allow in principle to construct higher Painlev\'e equations\footnote{See also \cite{DPNovikov},
\cite{Suleimanov12} and \cite{Tsuda}.} in terms of 2x2 linear problems related to spin chains via spectral duality
transformation. We are going to study this possibility in our future publications.

\section*{Appendix A: Special Cases} \addcontentsline{toc}{section}{Appendix A: Special Cases}
\def\theequation{A.\arabic{equation}}
\setcounter{equation}{0}

\subsection*{$b=(x-u(t))e^{g(t)x}$ and $b=(x-u(t))e^{g(t)x^2}$}

Let $b=(x-u(t))e^{g(t)x}$. The calculation similar to the one leading to (\ref{S403}) gives in this case:
 \beq\label{S403a} V_t(x)-V_t(u)-
\frac{V'(x)+V'(u)}{2(x-u)}+ \frac{V(x)\! -\! V(u)}{(x-u)^2}-\frac{\ddot g}{2}\, (x-u) -\frac{g}{2}\left (V'(x)\! -\!
V'(u)\right )=0
 \eeq
 and
 \beq\label{S403a01} H=\frac{1}{2}\left({\dot u}+\frac{g}{2}\right)^2+V(u,t)-\frac{1}{2}u{\dot g}+\frac{1}{8}g^2
 \eeq
with equation of motion
 \beq\label{S403a02} {\ddot u}=-V'(u)\,.
 \eeq

It is easy to see that the equation (\ref{S403a}) becomes  equivalent to (\ref{S403}) for the potential $\tilde V(\tilde x)$
after the change of variables
$$
x \to \tilde x = x-\frac{1}{2}\, G(t) \,, \quad \quad V(x)\to \tilde V(x)=V\Bigl (x-\frac{1}{2}\, G(t)\Bigr ) - \frac{\dot
g}{2}\, x\,,
$$
where $\dot G=g$. Notice also that the dependence $H({\dot u})$ in (\ref{S403a01}) can be obtained from (\ref{FG2033}) via
the local expansion (\ref{d21}). The later gives $b_1=e^{ug}$ and $b_2=g\, e^{ug}$. Then $v={\dot u}+\frac{g}{2}$.

Consider now the case $b=(x-u(t))e^{g(t)x^2}$. Let us perform the calculation similar to the one leading to (\ref{S403})
again. In this case we have:
 \beq\label{S403a10}
 \begin{array}{c}
 V_t(x)-V_t(u)-\frac{V'(x)+V'(u)}{2(x-u)}+ \frac{V(x)\! -\! V(u)}{(x-u)^2}-2{g}\left (V(x)\! -\! V(u)\right )-
\\ \ \\
-g\Big(x V'(x)+u V'(u)\Big) +(x^2-u^2)\left[ 3g{\dot g}-\frac{1}{2}\,{\ddot g}-2g^3 \right]=0
\end{array}
 \eeq
 and
 \beq\label{S403a11}
  H=\frac{1}{2}\left({\dot u}+{g}u\right)^2+V(u,t)+\frac{1}{2}(g^2-{\dot g})u^2+\frac{3}{2}g
 \eeq
with equation of motion
 \beq\label{S403a12} {\ddot u}=-V'(u)\,.
 \eeq
As in the previous example it can be shown that the equation (\ref{S403a10}) becomes  equivalent to (\ref{S403}) for the
potential $\tilde V(\tilde x)$ after the following change of variables:
 $$
x \to \tilde x = \alpha x=x\,e^{\int_t g(t)}\,,\ \alpha=e^{\int_t g(t)}\,,
 $$
 $$
V(x)\to \tilde V(x)=\alpha^2\Bigl( V(\alpha x)- x^2\left(g^2-\int_t \frac{{\ddot g} -2g{\dot g}}{2\alpha^2} \right) \Bigr )=
 $$
 $$
=e^{2\int_t g(t)}\left( V(x\,e^{\int_t g(t)})- x^2\left(g^2-\int_t\left[ e^{-2\int_t g(t)}(\frac{1}{2}{\ddot g} -g{\dot g})
\right]  \right) \right)\,.
$$

Notice also that the dependence $H({\dot u})$ in (\ref{S403a11}) can be obtained from (\ref{FG2033}) via the local expansion
(\ref{d21}). The later gives $b_1=e^{g u^2}$ and $b_2=2gu\, e^{g u^2}$. Then $v={\dot u}+gu$.

\subsection*{$b=(x-u_1(t))(x-u_2(t))(x-u_3(t))$}

When $b=(x-u_1)(x-u_2)(x-u_3)$ the coefficient behind the second order pole $\frac{1}{(x-u_1)^2}$ in (\ref{FG2}) have the
following form:
$$
\begin{array}{c}
V(x,t)-H+\frac{1}{2}{\dot u}_1^2+\frac{1}{2}\frac{{\dot u}_1+{\dot u}_2}{{ u}_1-{ u}_2}+\frac{1}{2}\frac{{\dot u}_1+{\dot
u}_3}{{u}_1-{u}_3}
-\frac{1}{2}\frac{1}{({u}_1-{ u}_2)^2}-\frac{1}{2}\frac{1}{({ u}_1-{ u}_3)^2}+\frac{1}{2}\frac{1}{({u}_1-{u}_2)({u}_1-{u}_3)}
\end{array}
$$
and two other coefficients can be obtained by the cyclic permutations. All three coefficients can not vanish simultaneously.
Therefore,  some other anzats for $W$ (\ref{W01}) should be used in this case. This notice reflects the fact  that
(\ref{W0})-(\ref{W01}) imply the one degree of freedom case.


\subsection*{$b=(x-u_1(t))^{\gamma}$ and $b=(x-u_1(t))^{\gamma_1}(x-u_2(t))^{\gamma_2}$}\label{sec_b}

Let us study the case $b=(x-u_1(t))^{\gamma}$, where $\gamma\in\CC^*$  (the case $\gamma=0$ is trivial). Notice that under
change $b\rightarrow b^\gamma$ the functions $f$ (\ref{FG1}) and $S$  (\ref{W1}) transform as follows:
 \beq\label{sb1}
 \begin{array}{c}
  f=\frac{b_x}{b}\longrightarrow \gamma\frac{b_x}{b}\,,\\
  S\longrightarrow V-H-\frac{1}{2}\gamma\frac{b_t}{b}+\frac{1}{2}\gamma\frac{b_{xx}}{b}+
  \frac{1}{2}\left(\frac{1}{4}\gamma^2-\gamma\right)\left(\frac{b_x}{b}\right)^2\,.
 \end{array}
 \eeq
For the case under consideration we have $f={\gamma}\frac{1}{x-u}$ and
 \beq\label{sb2}
 \begin{array}{c}
 S=V-H+\frac{\gamma}{2}{\dot u}\frac{1}{x-u}+\frac{1}{2}\left(\frac{1}{4}\gamma^2-\gamma\right)\frac{1}{(x-u)^2}\,.
 \end{array}
 \eeq
Substituting it into (\ref{FG2}) we obtain the following condition for cancellation of the forth and the third order poles:
 \beq\label{sb3}
 \begin{array}{l}
 (x-u)^{-4}:\ \ 0=\frac{1}{4}\gamma(\gamma^2-4\gamma+3)\,,\\
 (x-u)^{-3}:\ \  {\dot u}\left(\frac{1}{4}\gamma^2-\gamma\right)=-\frac{3}{4}\gamma^2 {\dot u}\,.
 \end{array}
 \eeq
The first one equation gives $\gamma=\{0,1,3\}$ while the second one $\gamma=\{0,1\}$. Therefore, the non-trivial solution is
 \beq\label{sb4}
\gamma=1\,.
 \eeq
Similarly, the case $b=(x-u_1(t))^{\gamma_1}(x-u_2(t))^{\gamma_2}$ leads to the following conditions:
 \beq\label{sb5}
 \begin{array}{l}
 (x-u_1)^{-4}: \frac{1}{4}\gamma_1(\gamma_1^2-4\gamma_1+3)=0\,,\\
 (x-u_1)^{-3}:\frac{1}{2}\frac{\gamma_1(\gamma_1-1)}{u_1-u_2}\left(2{\dot u}_1(u_1-u_2)+\gamma_2\right)\,,
\\ \ \\
 (x-u_2)^{-4}: \frac{1}{4}\gamma_2(\gamma_2^2-4\gamma_2+3)=0\,,\\
 (x-u_2)^{-3}: \frac{1}{2}\frac{\gamma_2(\gamma_2-1)}{u_2-u_1}\left(2{\dot u}_2(u_2-u_1)+\gamma_1\right)\,,
 \end{array}
 \eeq
which give
 \beq\label{sb7}
\gamma_1=\gamma_2=1\,.
 \eeq

\subsection*{$b=\exp\Big(\left(z/u(t)\right)^\gamma\Big)$}

First, it can be shown that $\gamma=0,1,2,3$...

Consider $\gamma=1$.  Substituting $b(z,u(t),t)=\exp(z/u(t))$ into (\ref{FG2}) we get:
 \beq\label{exot01}
 \begin{array}{c}
\left( -\frac{{\dot u}^2}{u^3}+\frac{1}{2}\frac{\ddot u}{u^2} \right)x+V_t-H_t-\frac{1}{2}\frac{{\dot
u}}{u^3}-\frac{1}{2u}{V'}_x=0\,.
 \end{array}
 \eeq
Applying $\p_x^2$ gives:
 \beq\label{exot02}
 \begin{array}{c}
V''_t-\frac{1}{2u}V'''=0\,.
 \end{array}
 \eeq
Notice that the function ${U}(z,{\dot u},{ u},t)$ satisfies the same equation even if we do not impose the condition
${U}=V(x,t)-H({\dot u},{ u},t)$. Under assumption ${U}=V(x,t)-H({\dot u},{ u},t)$ we have:
 \beq\label{exot03}
 \begin{array}{c}
V'''=V''_t=0\,.
 \end{array}
 \eeq
This leads to
 \beq\label{exot04}
 \begin{array}{c}
V(x,t)=\frac{\alpha}{2}x^2+b(t)x+c(t)\,,\ \ \alpha=\hbox{const}\,.
 \end{array}
 \eeq
Plugging it back to (\ref{exot01}) we obtain the following two equations (as coefficients behind $x^1$ and $x^0$):
 \beq\label{exot05}
 \left\{\begin{array}{l}
{\ddot u}=2\frac{{\dot u}^2}{u}-2{\dot b}u^2+\alpha u\,,\\ \ \\
H_t=\frac{1}{2}\frac{{\dot u}^2}{u^3}+{\dot c}-\frac{1}{2u}b\,.
 \end{array}\right.
 \eeq

\subsection*{Case 2 in (\ref{S706})}

Here it may be useful to use variable $u=u_1-\frac{1}{2}\sqrt{c-4t}$  (then $\dot u=\dot u_1+\frac{1}{\sqrt{c-4t}}$). Then
 \beq\label{S12} H=\frac{1}{2} (\dot u_1+\frac{1}{\sqrt{c-4t}})^2+V(u_1)=\frac{1}{2}\dot u^2+V(u+\frac{1}{2}\sqrt{c-4t}) \eeq
and, therefore,
 $$
\begin{array}{c}
V_t(x)-H_t -\frac{1}{2(x-u_1)}\Big(V'(x)-\ddot
u_1-2(c-4t)^{-\frac{3}{2}}-2\frac{V(x)-V(u_1)}{x-u_1}\Big) \\
\\ -\frac{1}{2(x-u_2)}\Big(V'(x)-\ddot
u_2+2(c-4t)^{-\frac{3}{2}}-2\frac{V(x)-V(u_2)}{x-u_2}\Big)=0
\end{array}
 $$
Cancellation of the first order poles at $x=u_{1,2}$ yields $ \ddot u_1=-V'(u_1)-2(c-4t)^{-\frac{3}{2}}$. On this equation
$H_t=V_t(u_1)-V'(u_1)\frac{1}{\sqrt{c-4t}}$. Thus we arrive to  \beq\label{S15}
\begin{array}{c}
V_t(x)-V_t(u_1)+V'(u_1)\frac{1}{\sqrt{c-4t}}-\frac{1}{2(x-u_1)}\Big(V'(x)+V'(u_1)-2\frac{V(x)-V(u_1)}{x-u_1}\Big) \\
\\ -\frac{1}{2(x-u_2)}\Big(V'(x)+V'(u_1)-2\frac{V(x)-V(u_2)}{x-u_2}\Big)=0
\end{array}\eeq
By analogy with   (\ref{S11}) we get

 \beq\label{S16} \left\{\begin{array}{l}
V_{t}^{\rm V}(x)=0 \\ \\
-20V_{t}^{\rm IV}(x)+V^{\rm VI}(x)=0 \\ \\
-13V^{\rm V}(x)+120V_{t}^{\rm III}(x)=0\\ \\
V^{\rm IV}(x)-6V_{t}^{\rm II}(x)=0\\ \\
6\p_z V_t(x)-V^{\rm III}(x)+\Big(-\frac{16}{13}t+\frac{4}{13}\Big)V_{t}^{\rm III}(x)=0
\end{array}\right.\eeq
From two upper equations it follows that $V(x)$ is the 6-th degree polynomial. Plugging it into (\ref{S16}) drops the degree
to 4 (similar to the Painlev{\'e} I, II cases). However, after substituting it back into (\ref{S15}) we get only trivial
solution
$$V(x,t)=f(t).$$

\section*{Appendix B: Elliptic Functions}
\addcontentsline{toc}{section}{Appendix B: Elliptic Functions}
\def\theequation{B.\arabic{equation}}
\setcounter{equation}{0}

Here we give a short version of the Appendix in \cite{ZZ12}.

\subsection*{Theta-functions.}
The Jacobi's theta-functions $\vartheta_a (z)= \vartheta_a (z|\tau )$, $a=0,1,2,3$, are defined by the formulas
 \beq\label{Bp1}
\begin{array}{l}
\vartheta _1(z)=-\displaystyle{\sum _{k\in \z}} \exp \left ( \pi i \tau (k+\frac{1}{2})^2 +2\pi i
(z+\frac{1}{2})(k+\frac{1}{2})\right ),
\\ \\
\vartheta _2(z)=\displaystyle{\sum _{k\in \z}} \exp \left ( \pi i \tau (k+\frac{1}{2})^2 +2\pi i z(k+\frac{1}{2})\right ),
\\ \\
\vartheta _3(z)=\displaystyle{\sum _{k\in \z}} \exp \left ( \pi i \tau k^2 +2\pi i z k \right ),
\\ \\
\vartheta _0(z)=\displaystyle{\sum _{k\in \z}} \exp \left ( \pi i \tau k^2 +2\pi i (z+\frac{1}{2})k\right ),
\end{array}
\eeq where $\tau$ is a complex parameter (the modular parameter) such that ${\rm Im}\, \tau >0$. Set
 $$
\omega _0 =0\,, \quad \omega_1 =\frac{1}{2}\,, \quad \omega _2=\frac{1+\tau}{2}\,, \quad \omega _3 =\frac{\tau}{2}\,,
 $$
then the function $\vartheta _a(z)$ has simple zeros at the points of the lattice $\omega _{a-1}+\ZZ +\ZZ \tau $ (here
$\omega_a \equiv \omega_{a+4}$).

\subsection*{Weierstrass $\wp$-function.}
The Weierstrass $\wp$-function is defined as
 \beq\label{Bp3} \wp (z)= -\p_z^2 \log \vartheta _1 (z)-2\eta \,, \eeq where
 \beq\label{Bp4} \eta = -\, \frac{1}{6}\, \frac{\vartheta _{1}^{'''}(0)}{\vartheta _1' (0)}= -\, \frac{2\pi i}{3}\, \p_{\tau}
\log \theta _1'(0|\tau ).
 \eeq
Its derivative is given by
 \beq\label{Bp5} \wp '(z)=-\, \frac{2\, (\vartheta _1'(0))^3}{\vartheta_2(0)\vartheta_3(0)
\vartheta_0(0)}\, \frac{\vartheta _2(z)\vartheta _3(z) \vartheta _0(z)}{\vartheta _{1}^{3}(z)}\,. \eeq

The values at the half-periods  \beq\label{Bp6} e_1 =\wp (\omega _1), \quad e_2 =\wp (\omega _2), \quad e_3 =\wp (\omega _3)
\eeq have special properties. For example, $e_1 +e_2 +e_3=0$. The differences $e_j-e_k$ can be represented in two different
ways:
 \beq\label{Bp7}
 \begin{array}{l}
\displaystyle{e_1 -e_2 =\pi ^2 \vartheta_0^4 (0)\, =\, 4\pi i \, \p_{\tau} \log\frac{\vartheta_3 (0)}{\vartheta_2 (0)}}
\\ \\
\displaystyle{e_1 -e_3 =\pi ^2 \vartheta_3^4 (0)\, =\, 4\pi i \, \p_{\tau} \log\frac{\vartheta_0 (0)}{\vartheta_2 (0)}}
\\ \\
\displaystyle{e_2 -e_3 =\pi ^2 \vartheta_2^4 (0)\, =\, 4\pi i \, \p_{\tau} \log\frac{\vartheta_0 (0)}{\vartheta_3 (0)}}\,.
 \end{array}
\eeq The second representation is a consequence of the heat equation (\ref{Bp2}) (see below):
 \beq\label{Bp8} e_k = 4\pi i
\, \p_{\tau}\Bigl ( \frac{1}{3}\, \log \vartheta _1'(0) -\log \vartheta_{k+1}(0)\Bigr ) \eeq
 or
 \beq\label{Bp9} \pi i \,
\p_{\tau}\log (e_j -e_k)=-e_l -2\eta \,, \eeq where $\{jkl\}$ - any cyclic permutation of $\{123\}$.
The $\wp$-function satisfies the differential equation  \beq\label{Bp10} (\wp '(z))^2 =4 (\wp (z)-e_1)(\wp (z)-e_2)(\wp
(z)-e_3). \eeq We also mention the formulae  \beq\label{Bp11a} \wp (z)-e_k =\frac{(\vartheta
_1'(0))^2}{\vartheta_{k+1}^2(0)}\, \frac{\vartheta_{k+1}^2(z)}{\vartheta_1^2(z)}\,. \eeq

\subsection*{Eisenstein functions and $\Phi$-function.}

By definition
 \beq\label{A.1} E_1(z)=\p_z\log\vth(z)\,, \quad \quad
E_2(z)=-\p_zE_1(z)= -\p_z^2\log\vth(z)=\wp(z)+2\eta \,. \eeq
 Behavior on the lattice:
  \beq\label{A.102}
E_1(z+1)=E_1(z),\ \ E_1(z+\tau)=E_1(z)-2\pi i\,,\eeq
  \beq\label{A.103}E_2(z+1)=E_2(z),\ \
E_2(z+\tau)=E_2(z)\,.\eeq
 The local expansion near $z=0$:
$$
E_1(z)=\frac{1}{z}-2\eta z  + \ldots \,,\quad \quad ~~E_2(z) =\frac{1}{z^2}+2\eta +\ldots
$$
Values at half-periods:  \beq\label{a21} E_1(\om_j)=-2\pi i\p_\tau\om_j \eeq and, therefore,  \beq\label{a22}
E_1(\om_j)+E_1(\om_k)=E_1(\om_j+\om_k) \eeq holds true for any different $j,k =1,2,3$.

Another useful function is  \beq\label{Bp11} \Phi(u,z)= \frac{\vth(u+z)\vth'(0)}{\vth(u)\vth(z)}\,. \eeq

It has the following properties: $$\Phi(u,z)=\Phi(z,u)\,,$$ $$\Phi(-u,-z)=-\Phi(u,z)\,,$$
 \beq\label{Bp1101}
\Phi(u,z)\Phi(-u,z)=\wp(z)-\wp(u) \eeq
 \beq\label{Bp1102} \Phi(u,z)\Phi(w,z)=\Phi(u+w,z)(E_1(z)+E_1(u)+E_1(w)-E_1(z+u+w)). \eeq
 \beq\label{A.3a}
\Phi(u,z)=\frac{1}{z}+E_1(u)+\frac{z}{2}(E_1^2(u)-\wp(u))+ O(z^2). \eeq
 \beq\label{A3b} \p_z \Phi (u,z)=\Phi(u,z)
(E_1(u+z)-E_1(z)). \eeq

Behavior on the lattice:  \beq\label{A.14} \Phi(u,z+1)=\Phi(u,z)\,,~~~\Phi(u,z+\tau)=e^{-2\pi i u}\Phi(u,z)\,. \eeq

Is is also convenient to introduce
 \beq\label{Bp23} \vf_j(z)=e^{2\pi i z\p_\tau \om_j}\Phi(z,\om_j)\,, \quad j=1,2,3
\eeq with properties:
 \beq\label{Bp24}
\vf_j^2(z)=\wp(z)-e_j,\ \ \ \vf_j^2(z)-\vf_k^2(z)=e_k-e_j \eeq
 \beq\label{Bp25}
\vf_j(z)\vf_k(z)=\vf_l(z)(E_1(z)+E_1(\om_l)-E_1(z+\om_l)).
 \eeq
 \beq\label{Bp26} \p_z\vf_j(z)=\vf_j (z)\Bigl [ E_1(z+\om_j )-E_1(\om_j)-E_1(z)\Bigr ]=-\vf_k(z)\vf_l(z), \eeq where
$j,k,l$ is any cyclic permutation of $1,2,3$.

\subsection*{Heat equation and related formulae}

All the theta-functions satisfy the ``heat equation''
 \beq\label{Bp2} 4\pi i \p_{\tau}\vartheta _a(z|\tau )= \p_{z}^{2}\vartheta _a(z|\tau ) \eeq or
 $$2\p_{t}\vartheta _a(z )=\p_{z}^{2}\vartheta _a(z)\ \ \ \displaystyle{t=\frac{\tau}{2\pi i}}\,.$$

One can also introduce the ``heat coefficient'' $\displaystyle{\kappa=\frac{1}{2\pi i}}$ and rewrite the heat equation in the
form $\displaystyle{\p_{\tau}\vartheta _a(z|\tau )= \frac{\kappa}{2}\, \p_{z}^{2}\vartheta _a(z|\tau )}$. All formulas for
derivatives of elliptic functions with respect to the modular parameter are based on the heat equation.

The $\tau$-derivatives are given by the following
\begin{predl}
The identities  \beq\label{Bp12} \p_\tau\Phi(z,u)=\kappa\p_z\p_u\Phi(z,u), \eeq  \beq\label{Bp13} \p_\tau
E_1(z)=\frac{\kappa}{2}\, \p_z(E_1^2(z)-\wp(z)), \eeq  \beq\label{Bp14} \p_\tau E_2(z)=\kappa E_1(z)E_2'(z)-\kappa
E_2^2(z)+\frac{\kappa}{2}\, \wp''(z), \eeq with the ``heat coefficient'' $\displaystyle{\kappa=\frac{1}{2\pi i}}$, hold
true\footnote{(\ref{Bp12}) was obtained in \cite{LO97},\cite{Takasaki02}.}.
\end{predl}
The proof can be found in \cite{ZZ12}.

Introduce now
 \beq\label{6p1}
X(x,t)=\frac{\wp (x)-e_1}{e_2 -e_1}\,, \quad \quad T(t)=\frac{e_3 -e_1}{e_2 -e_1}=\left ( \frac{\vartheta _3(0|\tau
)}{\vartheta _0 (0|\tau )}\right )^4.
 \eeq

Then we have
 \beq\label{6p4} X=\frac{\wp (x)-e_1}{e_2 -e_1}\,, \quad \quad X\! - \! 1 =\frac{\wp (x)-e_2}{e_2
-e_1}\,, \quad \quad X\! - \! T =\frac{\wp (x)-e_3}{e_2 -e_1}\,,
 \eeq
and, therefore,
 \beq\label{6p6} \left (\frac{\p X}{\p x}\right )^2 = 4(e_2 -e_1)\, X(X-1)(X-T) \eeq
  \beq\label{6p5}
\frac{\p ^2 X}{\p x^2} = 2(e_2 -e_1)\, X(X-1)(X-T)\left ( \frac{1}{X}+ \frac{1}{X-1}+\frac{1}{X-T}\right )\,.
 \eeq
Let us give some more relations:
 \beq\label{6p7}
\begin{array}{rll}
\displaystyle{\frac{(e_2 -e_1)T}{X}}&=& \wp (x+\omega _1) -e_1\,,
\\ &&\\
\displaystyle{-\, \frac{(e_2 -e_1)(T-1)}{X-1}}&=& \wp (x+\omega _2) -e_2\,,
\\ &&\\
\displaystyle{\frac{(e_2 -e_1)T(T-1)}{X-T}}&=& \wp (x+\omega _3) -e_3\,,
\end{array}
 \eeq
 \beq\label{6p8}
\frac{\p T}{\p t}=2(e_2 -e_1) T(T-1)\,. \eeq
 \beq\label{6p88}
\p_T(e_2-e_1)=\p_t(e_2-e_1)\frac{1}{T_t}=-\frac{e_3+2\eta_1}{T(T-1)}\,. \eeq
The following identity holds true\footnote{This formula was proved by K.Takasaki in \cite{Takasaki01} by comparison of
analytic properties of the both sides. In \cite{ZZ12} the proof was given by a direct computation.}:
 \beq\label{6p9} \!\!\! \frac{\p
X}{\p t}=\frac{\p X}{\p x}\, \, \frac{\vartheta _0'(x)}{\vartheta _0(x)}
 \eeq
 or
 \beq\label{Bp27} \p_\tau X
 =\kappa \p_z X\left(E_1(z+\om_3)-E_1(\om_3)\right)=\kappa \, \p_z X\, \p_z \log \theta_0(z)\,.
 \eeq

\section*{Appendix C: ${\bf U}$-${\bf V}$ pairs for ${\rm P}_{\rm I}$-${\rm P}_{\rm V}$ }
\addcontentsline{toc}{section}{Appendix C: ${\bf U}$-${\bf V}$ pairs for ${\rm P}_{\rm I}$-${\rm P}_{\rm V}$}
\def\theequation{C.\arabic{equation}}
\setcounter{equation}{0}

Here we list the ${\bf U}$-${\bf V}$ pairs for ${\rm P}_{\rm I}$-${\rm P}_{\rm V}$ satisfying zero curvature equation
(\ref{P1c}) and admitting the quantum Painlev\'e-Calogero correspondence. The ${\rm P}_{\rm VI}$ case is too complicated. In
principle, it is gauge equivalent to different types of known elliptic $2\times 2$ ${\bf U}$-${\bf V}$ pairs (see
\cite{Zotov04},\cite{LOZ2}) which are in their turn related by Hecke transformations \cite{LOZ1,LOSZ}.

\paragraph{\underline{Painlev{\'e} I}}
 \beq\label{P1} 4\ddot u=6u^2 +t\,, \eeq
 \beq\label{P1a}
H_{\rm I}(p, u)=\frac{p^2}{2} -\frac{u^3}{2} -\frac{tu}{4}\,, \eeq
\beq\label{matricesU} {\bf U}(x,t)= \left (
\begin{array}{cc}
\dot u & x-u\\ &\\
x^2 \! +\! xu \! +\! u^2 \! +\! \frac{1}{2}\,t &  -\dot u
\end{array}\right ),\quad \quad
{\bf V} (x,t) =\left ( \begin{array}{cc}
0& \frac{1}{2}\\ &\\
\frac{1}{2}\, x +u & 0
\end{array}\right )\,.
\eeq

\paragraph{\underline{Painlev{\'e} II}}
\beq\label{P2} \ddot u=2u^3 +tu-\alpha \,, \eeq
\beq\label{P2aa}
\begin{array}{c}
H_{\rm II}(p, u)= \displaystyle{\frac{p^2}{2} -\frac{1}{2} \left (u^2+\frac{t}{2}\right )^2 +\alpha u}\,,
\end{array}
\eeq
 \beq\label{matricesU21}
{\bf U}= \left (
\begin{array}{cc}
x^2+\dot u -u^2 & x-u\\ &\\
(x+u)(2u^2 \! -\! 2\dot u \! +\! t)\! -\! 2\alpha \! - \! 1 &  -x^2\! -\! \dot u \!  +\! u^2
\end{array}\right )\,,
 \eeq
 \beq\label{matricesU22}
{\bf V}  =\left (
\begin{array}{cc}
\frac{x+u}{2}& \frac{1}{2}\\ &\\
u^2 \! - \! \dot u \! +\! \frac{t}{2} & -\, \frac{x+u}{2}\end{array}\right )\,.
 \eeq

\paragraph{\underline{Painlev{\'e} III}}
 \beq\label{P3ak}\begin{array}{l} 2\ddot u=e^t(\alpha e^{2u} +\beta e^{-2u}) + e^{2t}(\gamma e^{4u} +\delta e^{-4u}).
 \end{array}\eeq
 \beq\label{P3aak}
 \begin{array}{l}
H_{\rm III}(p, u)= \displaystyle{\frac{p^2}{2} -\nu^2 e^t \cosh (2u-2\varrho ) -\mu^2 e^{2t}\cosh (4u).}
\end{array}
 \eeq
 \beq\label{la1}\begin{array}{l} {\bf U}_{11}=\dot{u}e^{2u-2x}+\theta\Big(1-e^{2u-2x} \Big)+\frac{1}{2}\Big(
e^{2x+t}-e^{2u-2x}-e^{4u+t-2x}+1\Big)\,,\ \ \ \ \ \ \ \ \ \ \ \ \ \
 \end{array}
 \eeq
 \beq\label{la2}\begin{array}{l}
 {\bf U}_{12}= e^{\frac{t}{2}}\Big(e^{-u+x}-e^{u-x}\Big)\,,
 \hskip105mm
 \end{array}
 \eeq
  \beq\label{la3}\begin{array}{l}
{\bf U}_{21}=\dot{u}^2 e^{u-\frac{t}{2}-3x}\Big(e^{2x}+e^{2u}\Big)
-\dot{u}e^{u-\frac{t}{2}-3x}\Big(e^{2x}+e^{2u+t+2x}+(1+2\theta)e^{2u}
+e^{4u+t}\Big)\\ \ \\
+ \theta^2\left( -e^{u-\frac{t}{2}-x}+e^{3u-\frac{t}{2}-3x} \right)+\theta\left(
e^{3u-\frac{t}{2}-3x}+e^{5u+\frac{t}{2}-3x}\right)+4\lambda
e^{-u+\frac{t}{2}-x}\\ \ \\
- 4\chi\left(e^{-3u+\frac{3t}{2}-x}+e^{-u+\frac{3t}{2}-3x} \right)
\\ \ \\
+ \frac{1}{4}\Big( e^{u-\frac{t}{2}-x}
+2e^{3u+\frac{t}{2}-x}+e^{5u+\frac{3t}{2}-x}+e^{3u-\frac{t}{2}-3x}+2e^{5u+\frac{t}{2}-3x}+e^{7u+\frac{3t}{2}-3x}\Big)\,.
\end{array}
\eeq

 \beq\label{la4}\begin{array}{l} {\bf V}_{11}=-\frac{1}{2}\dot{u}\Big(e^{2u\!-\!2x}\!+\!1\Big)+\frac{\theta}{2}\Big(1\!+\!e^{2u\!-\!2x}
\Big)
\!+\!\frac{1}{4}\Big( e^{2x\!+\!t}\!+\!e^{2u\!-\!2x}\!+\!e^{4u\!+\!t\!-\!2x}\!+\!1\!+\!2e^{2u\!+\!t}\Big)\,,\hskip14mm
 \end{array}
 \eeq
  \beq\label{la5}\begin{array}{l}
{\bf V}_{12}= \frac{1}{2}e^{\frac{t}{2}}\Big(e^{-u+x}+e^{u-x}\Big)\,,\hskip90mm
 \end{array}
 \eeq
 \beq\label{la6}\begin{array}{l}
{\bf V}_{21}=\frac{1}{2}\dot{u}^2 e^{u-\frac{t}{2}-3x}\Big(e^{2x}\!-\!e^{2u}\Big)
-\frac{1}{2}\dot{u}e^{u-\frac{t}{2}-3x}\Big(e^{2x}\!+\!e^{2u+t+2x}\!-\!(1+2\theta)e^{2u} \!-\!e^{4u+t}\Big)
\\ \ \\
- \frac{\theta^2}{2}\left( e^{u-\frac{t}{2}-x}+e^{3u-\frac{t}{2}-3x} \right)-\frac{\theta}{2}\left(
e^{3u-\frac{t}{2}-3x}+e^{5u+\frac{t}{2}-3x}\right)+2\lambda
e^{-u+\frac{t}{2}-x}\\ \ \\
- 2\chi\left(e^{-3u+\frac{3t}{2}-x}-e^{-u+\frac{3t}{2}-3x} \right)
\\ \ \\
+ \frac{1}{8}\Big( e^{u-\frac{t}{2}-x}
+2e^{3u+\frac{t}{2}-x}+e^{5u+\frac{3t}{2}-x}-e^{3u-\frac{t}{2}-3x}-2e^{5u+\frac{t}{2}-3x}-e^{7u+\frac{3t}{2}-3x}\Big)\,.
\end{array}
\eeq

Notice, that an interesting equation holds:
 \beq\label{la7} \p_x\Big({\bf U}_{21}e^{2x} \Big)=2 \Big({\bf V}_{21}e^{2x} \Big)
 \eeq
(in this case $X=e^{2x}$). Therefore, some relation exists between ${\bf U}_{21}$ and ${\bf V}_{21}$ elements just as for
(12)-elements. For example, for ${\rm P}_{\rm II}$ we have $\p_x {\bf U}_{21}=2 {\bf V}_{21}$.

\paragraph{\underline{Truncated Painlev{\'e} III \cite{AmAr}:}} $\ddot u =2\nu^2 e^t\sinh (2u )$
 \beq\label{matricesU3a} {\bf U}(x,t) = \left (
\begin{array}{cc}
\dot u &2\nu e^{t/2}\sinh (x-u )
\\ & \\
2\nu e^{t/2}\sinh (x+u )  & -\dot u
\end{array}\right )\,,
 \eeq
 \beq\label{matricesU3b} {\bf V }(x,t) =\left (
\begin{array}{cc}
0& \nu e^{t/2}\cosh (x-u )
\\ & \\
\nu e^{t/2} \cosh (x+u )& 0
\end{array}\right )\,.
 \eeq

\paragraph{\underline{Painlev{\'e} IV}}
\beq\label{P4a} \ddot u=\frac{3}{4}\, u^5 +2tu^3 + (t^2 -\alpha ) u +\frac{\beta}{2u^3}\,, \eeq
\beq\label{P4aa}
\begin{array}{lll}
H_{\rm IV}^{(\alpha , \beta )}(p, u)&=& \displaystyle{\frac{p^2}{2} -\frac{u^6}{8} -\frac{tu^4}{2} -\frac{1}{2}\left (t^2
-\alpha \right )u^2 +\frac{\beta}{4u^2}\,.}
\end{array}
\eeq
\beq\label{matricesU4a} {\bf U}= \left (
\begin{array}{cc}\displaystyle{
\frac{x^3}{2}\! +\! tx \! +\! \frac{Q+\frac{1}{2}}{x}}&x^2 -u^2
\\ & \\
\displaystyle{\frac{Q^2+\frac{\beta}{2}}{u^2x^2}\! -\! Q\! -\! \alpha \! -\! 1} & \,\,\,\,\displaystyle{ -\frac{x^3}{2}\! -\!
tx \! -\! \frac{Q+\frac{1}{2}}{x}}
\end{array}\right )\,,
\eeq

\beq\label{matricesU4b} {\bf V} =\left (
\begin{array}{cc}
\displaystyle{ \frac{x^2+u^2}{2} + t}&\,\, x
\\ & \\
\displaystyle{ -\, \frac{Q+\alpha +1}{x}}&\,\,\,\,\,\,\,\, \displaystyle{ -\frac{x^2 +u^2}{2} - t}
\end{array}\right )\,,
\eeq where
$$
Q=u\dot u -\frac{u^4}{2}-tu^2.
$$

\paragraph{\underline{Painlev{\'e} V}}
\beq\label{P5aaa4}\begin{array}{c} \ddot u =-\frac{2\alpha \cosh u}{\sinh^3 u}- \frac{2\beta \sinh u}{\cosh^3 u} -\gamma
e^{2t}\sinh (2u) -\frac{1}{2}\, \delta e^{4t}\sinh (4u)\,,\end{array} \eeq
\beq\label{P5aaab4}\begin{array}{c} H_{\rm V}(p, u) = {\frac{p^2}{2}-\, \frac{\alpha}{\sinh^2 x} -\, \frac{\beta}{\cosh^2
x}+\frac{\gamma e^{2t}}{2} \cosh (2x)+\frac{\delta e^{4t}}{8}\cosh (4x)}\,, \end{array}\eeq

 \beq\label{la8}\begin{array}{l} {\bf U}_{11}= \dot{u}\frac{\sinh (2u)}{\sinh (2x)}-\frac{2\sigma}{\sinh(2x)}\Big(
\cosh(2x)-\cosh(2u)\Big)\hskip50mm
\\ \ \\
+\frac{e^{2t}}{4\sinh(2x)}\Big( \cosh(4x)-\cosh(4u)\Big)+\coth(2x)\,.
 \end{array}
 \eeq
 \beq\label{la9}\begin{array}{l}
{\bf U}_{12}= e^t\Big( \cosh(2x)-\cosh(2u)\Big)\,.\hskip77mm
 \end{array}
 \eeq
 \beq\label{la10}\begin{array}{l}
{\bf U}_{21}= \dot{u}^2\frac{e^{-t}}{\sinh^2(2x)}\Big( \cosh(2u)+\cosh(2x)\Big)
\\ \ \\
+\dot{u}\frac{\sinh(2u)}{\sinh^2(2x)}\Big( 4\sigma e^{-t}-e^t \Big[\cosh(2u)+\cosh(2x)\Big]
\Big)\\ \ \\
+8\sigma^2e^{-t}\frac{\coth^2(u)}{\sinh^2(2x)}\Big( \sinh^2(u)-\cosh^2(x)\Big)-2\sigma
e^t\frac{\sinh^2(2u)}{\sinh^2(2x)}\\ \ \\
-2e^{-t}\frac{\xi^2+2\xi\sigma}{\sinh^2(u)\sinh^2(x)}+2e^{-t}\frac{\zeta^2}{\cosh^2(u)\cosh^2(x)}
+ \frac{e^{3t}\sinh^2(2u)}{4\sinh^2(2x)}\Big( \cosh(2u)+\cosh(2x)\Big)\,.
 \end{array}
 \eeq

\beq\label{la11}\begin{array}{l} {\bf V}_{11}= \frac{1}{2}e^{2t}\Big( \cosh(2x)+\cosh(2u)\Big)-2\sigma
+\frac{1}{2}\,,\hskip54mm
\\ \ \\
{\bf V}_{12}= e^t\sinh(2x)\,,
\\ \ \\
{\bf V}_{21}=\frac{e^{-t}}{\sinh(2x)}\Big(
\Big(\dot{u}^2\!-\!\frac{1}{2}\dot{u}e^{2t}\sinh(2u)\Big)^2\!+\!\frac{4\zeta^2}{\cosh^2(u)}
\!-\!4\frac{\xi^2\!+\!2\xi\sigma}{\sinh^2(u)}\!-\!4\sigma^2\coth^2(u)\Big)\,.
\end{array}
\eeq

\subsubsection*{Acknowledgments}

\addcontentsline{toc}{section}{Acknowledgments}

We are grateful to A.Morozov for discussions. The work was supported in part by Ministry of Science and Education of Russian
Federation under contract 8207. The work of A.Zabrodin was also supported in part by RFBR grant 11-02-01220, by joint RFBR
grants 12-02-91052-CNRS, 12-02-92108-JSPS, by grant NSh-3349.2012.2 for support of leading scientific schools. The work of
A.Zotov was also supported in part by RFBR-12-01-00482, RFBR-12-01-33071 mol$\_$a$\_$ved, by joint RFBR-13-02-90470
Ukr$\_$f$\_$a, by the Russian President fund MK-1646.2011.1 and by the "Dynasty" fund.

\begin{small}

\end{small}
\end{document}